\documentclass[conference]{IEEEtran}
\IEEEoverridecommandlockouts
\usepackage{cite}
\usepackage{amsmath,amssymb,amsfonts}
\usepackage{graphicx}
\usepackage{textcomp}
\usepackage{xcolor}
\def\BibTeX{{\rm B\kern-.05em{\sc i\kern-.025em b}\kern-.08em
    T\kern-.1667em\lower.7ex\hbox{E}\kern-.125emX}}

\usepackage{multirow}
\usepackage{listings}
\usepackage{graphicx}
\usepackage{url}
\usepackage{algorithm}
\usepackage{subfigure}
\usepackage{float}
\usepackage{enumitem}

\usepackage{hyphenat}
\newfloat{algorithm}{t}{lop}
\usepackage[noend]{algpseudocode}

\newcommand{\cb}{}

\def\vsp{0.05in}

\def\sysname{G-TADOC}
\def\avgSpeedup{31.1}

\def\initTimeSaving{76.5}

\def\traversalTimeSaving{82.2}

\begin{document}

\bstctlcite{IEEEexample:BSTcontrol}

\title{
G-TADOC: Enabling Efficient GPU-Based\\ Text Analytics without Decompression
}

\author{\IEEEauthorblockN{Feng Zhang\IEEEauthorrefmark{1},
Zaifeng Pan\IEEEauthorrefmark{6},
Yanliang Zhou\IEEEauthorrefmark{1},
Jidong Zhai\IEEEauthorrefmark{2}, Xipeng Shen\IEEEauthorrefmark{3},
Onur Mutlu\IEEEauthorrefmark{4}, Xiaoyong Du\IEEEauthorrefmark{1}}
\IEEEauthorblockA{\IEEEauthorrefmark{1}\resizebox{1\textwidth}{!}{Key Laboratory of Data Engineering and Knowledge Engineering (MOE), and School of Information, Renmin University of China}\\
\IEEEauthorrefmark{6}School of Mechanical Engineering, Shanghai Jiao Tong University	\\
\IEEEauthorrefmark{2}Department of Computer Science and Technology, Tsinghua University, BNRist\\
\IEEEauthorrefmark{3}Computer Science Department, North Carolina State University\\
\IEEEauthorrefmark{4}Department of Computer Science, ETH Z{\"u}rich\\
fengzhang@ruc.edu.cn,
764556762@sjtu.edu.cn,
triode-zyl@ruc.edu.cn,
zhaijidong@tsinghua.edu.cn,\\
xshen5@ncsu.edu,
onur.mutlu@inf.ethz,
duyong@ruc.edu.cn
}}

\maketitle

\bstctlcite{IEEEexample:BSTcontrol}

\begin{abstract}

Text analytics directly on compression (TADOC) has proven to be a promising technology for big data analytics. 
GPUs are extremely popular accelerators for data analytics systems. Unfortunately, no work so far  
shows how to utilize GPUs to accelerate TADOC. We describe G-TADOC, the first framework that provides 
GPU-based text analytics directly on compression, effectively enabling efficient text analytics on GPUs 
without decompressing the input data.

G-TADOC solves three major challenges. First, TADOC involves
a large amount of dependencies, which makes it difficult to exploit massive parallelism on a GPU. 
We develop a novel fine-grained thread-level workload scheduling strategy for GPU threads, which partitions heavily-dependent loads adaptively in a fine-grained manner. 
Second, in developing G-TADOC, thousands of GPU threads writing to the same result
buffer leads to inconsistency while directly using locks and atomic operations lead to large synchronization overheads. We
develop a memory pool with thread-safe data structures on GPUs to handle such difficulties. Third, maintaining the sequence
information among words is essential for lossless compression. We design a sequence-support strategy, which maintains
high GPU parallelism while ensuring sequence information. 

Our experimental evaluations show that G-TADOC provides \avgSpeedup{}$\times$ average speedup compared to state-of-the-art TADOC.

\end{abstract}


\section{Introduction}

Text analytics directly on compression (TADOC)~\cite{Zhang2018ics,zhang2018efficient,zhang2020enabling,zhang2020tadoc} has proven to be a promising technology for big data analytics.
Since TADOC processes compressed data without decompression,
a large amount of space can be saved.
Meanwhile,  TADOC reuses both data and intermediate computation results,
which results in that the same contents in different parts of original files can be processed only once, thus saving significant computation time.
Recent studies show that TADOC can save up to half of the processing time and 90.8\% storage space~\cite{Zhang2018ics,zhang2018efficient}.
On the other hand, GPU as a heterogeneous processor shows promising performance in many real applications, such as artificial intelligence. 
It is popular to use heterogeneous processors, such as GPU, to accelerate data  analytics systems~\cite{root2016mapd,yuan2016spark,koliousis2016saber,wang2017gunrock,zhang2020finestream}.
Therefore, it is essential to enable efficient data analytics on GPUs without decompression.

Applying GPUs to accelerate TADOC brings three key benefits.
First,
GPU performance is much higher than CPU performance,
so applying GPUs with proper designs can greatly accelerate TADOC performance,
which means that users can feel no delay in data analytics towards massive data.
Second,
previous TADOC mainly focuses on distributed systems.
If we develop a GPU-based solution on a single HPC server while achieving higher performance,
tremendous resources, including equipment cost and electricity cost, can be significantly saved.
Third,
many data analytics applications, such as {\em latent Dirichlet allocation} (LDA)~\cite{blei2003latent} and \emph{term frequency-inverse document frequency} (TFIDF)~\cite{joachims1996probabilistic}, have been ported to GPUs,
while TADOC has proven to be suitable for these advanced data analytics applications.
Hence, providing a GPU solution would remove the last barrier to apply TADOC to a wide range of applications.

Although it is both beneficial and essential to develop TADOC on GPUs,
building efficient GPU-based TADOC is very challenging.
Applying GPUs to accelerate TADOC faces three challenges.
First,
TADOC organizes data into rules, which can further be represented as a DAG.
Unfortunately, the amount of dependencies among the rule-structured DAG of TADOC is extremely large, which is unfriendly for GPU parallelism.
For example, in our experiments,
the generated DAG for each file has 450,704 dependent middle-layer nodes on average,
which greatly limits its parallelism.
Even worse,
a node in the  DAG of TADOC can have multiple parents,
which makes this problem more complicated.
Second, a large number of GPU threads writing to the same result buffer  inevitably cause tremendous write conflicts.
A straightforward solution is to lock the buffer for threads, but such atomicities lose partial performance.
In the worst case,
the parallel performance is lower than that of the CPU sequential TADOC.
Third,
maintaining and utilizing the sequence information on GPUs is another difficulty:
the original TADOC adopts  a recursive call to complete sequential traversal on compressed data,
which is similar to a depth-first search (DFS) and is extremely hard to solve in parallel.

Currently, none of the TADOC solutions can solve the challenges of enabling TADOC on GPUs mentioned above.
Zhang \emph{et al}.~\cite{zhang2018efficient} first proposed TADOC solution but it is designed in a sequential manner.
Although TADOC can be applied in a distributed environment,
TADOC adopts coarse-grained parallelism and the processing for each compressed unit is still sequential.
Zhang \emph{et al}. next developed a domain specific language (DSL), called Zwift, to present TADOC~\cite{Zhang2018ics},
and further realized random accesses on compressed data~\cite{zhang2020enabling}.
However, the parallelism problems still exist.
Zhang \emph{et al}.~\cite{zhang2020tadoc} then provided a parallel TADOC design,
which provides much better performance than the sequential TADOC.
Unfortunately, such parallelism is still coarse-grained: it only divides the original file into several sub-files, processes different files separately, and then follows a merge process,
which cannot be utilized by GPUs efficiently.

We design  G-TADOC, the first framework that provides 
\textbf{\underline{G}}PU-based \textbf{\underline{t}}ext \textbf{\underline{a}}nalytics \textbf{\underline{d}}irectly \textbf{\underline{o}}n \textbf{\underline{c}}ompression, effectively enabling efficient text analytics on GPUs 
without decompressing input data. 
\sysname{} involves three novel features that can address the above three challenges.
First,
to utilize the GPU parallelism,
we develop a fine-grained thread-level  workload  scheduling strategy on GPUs,
which allocates thread resources according to the load of different rules adaptively and uses masks to describe the relations between rules
(Section~\ref{subsec:paraExe}).
Second,
to solve the challenge of write conflict from multiple threads,
we enable \sysname{} to maintain its own memory pool and design thread-safe data structures.
We use a lock buffer when multiple threads update the global results simultaneously
(Section~\ref{subsec:dataStructure}).
Third,
to support sequence sensitive applications in \sysname{},
we develop \emph{head} and \emph{tail} data structures in each rule to store the contents at the beginning and end of the rule,
which requires a light-weight DAG traversal (detailed in Section~\ref{subsec:seqSupport}).

We evaluate \sysname{} on three GPU platforms, which involve three generations of Nvidia GPUs (Pascal, Volta, and Turing micro-architectures),
and use five real-world datasets of varying lengths, structures, and content.
Compared to  TADOC on CPUs,
\sysname{} achieves \avgSpeedup{}$\times$ speedup.
In detail,
TADOC can be divided into two phases:
initialization and DAG traversal.
For the initialization phase, \sysname{} achieves \initTimeSaving{}\% time saving,
while for the DAG traversal phase, \sysname{} achieves  \traversalTimeSaving{}\% time saving.

As far as we know,
this is the first work enabling efficient text analytics on GPU without decompression.
In summary,
we have made the following contributions in this work.

\vspace{0.05in}

\begin{itemize}

\item
We present \sysname{},
which is the first framework enabling efficient GPU-based text analytics directly on compressed data.

\item
We unveil the challenges for developing TADOC on GPUs and provide a set of solutions to these challenges.

\item
We evaluate \sysname{} on three GPU platforms, and demonstrate its significant benefits compared to  state-of-the-art TADOC.

\end{itemize}

\section{Background}
In this section,
we introduce TADOC and GPUs,
which are the background and premises of our work.

\subsection{TADOC}
TADOC~\cite{Zhang2018ics,zhang2018efficient,zhang2020enabling,zhang2020tadoc} is a novel lossless compression technique that enables data analytics directly on compressed data without decompression.
In detail,
TADOC adopts dictionary conversion to encode original input data with numbers,
and then uses context-free grammar (CFG) to recursively represent the numerical transformed data after conversion into rules.
Repeated pieces of data  are transformed into different rules in CFG,
and the data analytics tasks are then represented as rule interpretations.
To leverage redundant information between files,
TADOC inserts  unique splitting symbols for file boundaries.
 Moreover, the CFG can be represented as a directed acyclic graph (DAG),
 so the interpretation of the rules for data analytics can be regarded as a DAG traversal problem.
Currently, TADOC extends Sequitur~\cite{nevill1996inferring,nevill1997identifying,nevill1997linear} as its core algorithm.

\begin{figure}[h!]

\includegraphics[width=1\linewidth]{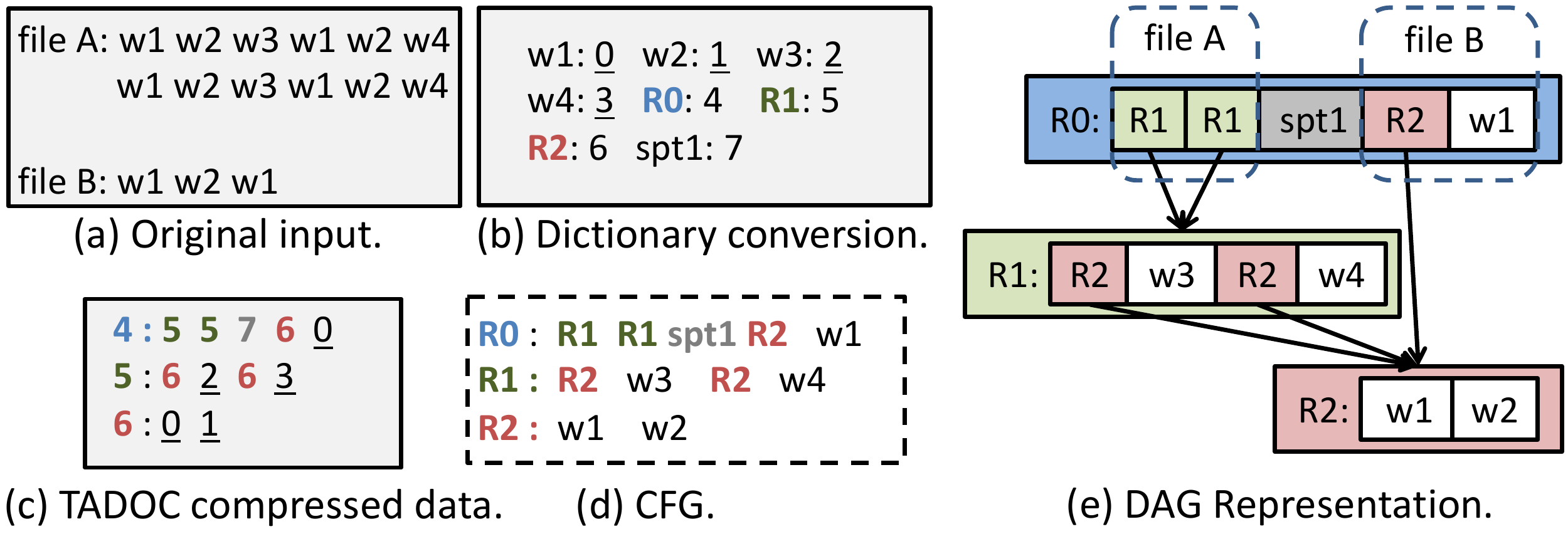}

  \caption{A compression example with TADOC.}
  \label{fig:SequiturExample}

\end{figure}

We use Figure~\ref{fig:SequiturExample} to show how TADOC compresses data by CFG representation, 
which is an example used in~\cite{zhang2018efficient}.
Figure~\ref{fig:SequiturExample}~(a) shows the original input data, which consists of two files: file \emph{A} and file \emph{B}, and
``wi'' represents a unique word.
Figure~\ref{fig:SequiturExample}~(b) shows the dictionary conversion,
which uses an integer to represent an element.
Note that the rules ``Ri'' and file splitters ``spti'' are also transformed into numerical forms.
Figure~\ref{fig:SequiturExample}~(c) shows the TADOC compressed data,
which are sequences of numbers.
The TADOC compressed data can be viewed as CFG shown in Figure~\ref{fig:SequiturExample}~(d), which can be further organized as a DAG shown in Figure~\ref{fig:SequiturExample}~(e) for traversals.

\begin{figure}[h!]
\includegraphics[width=1\linewidth]{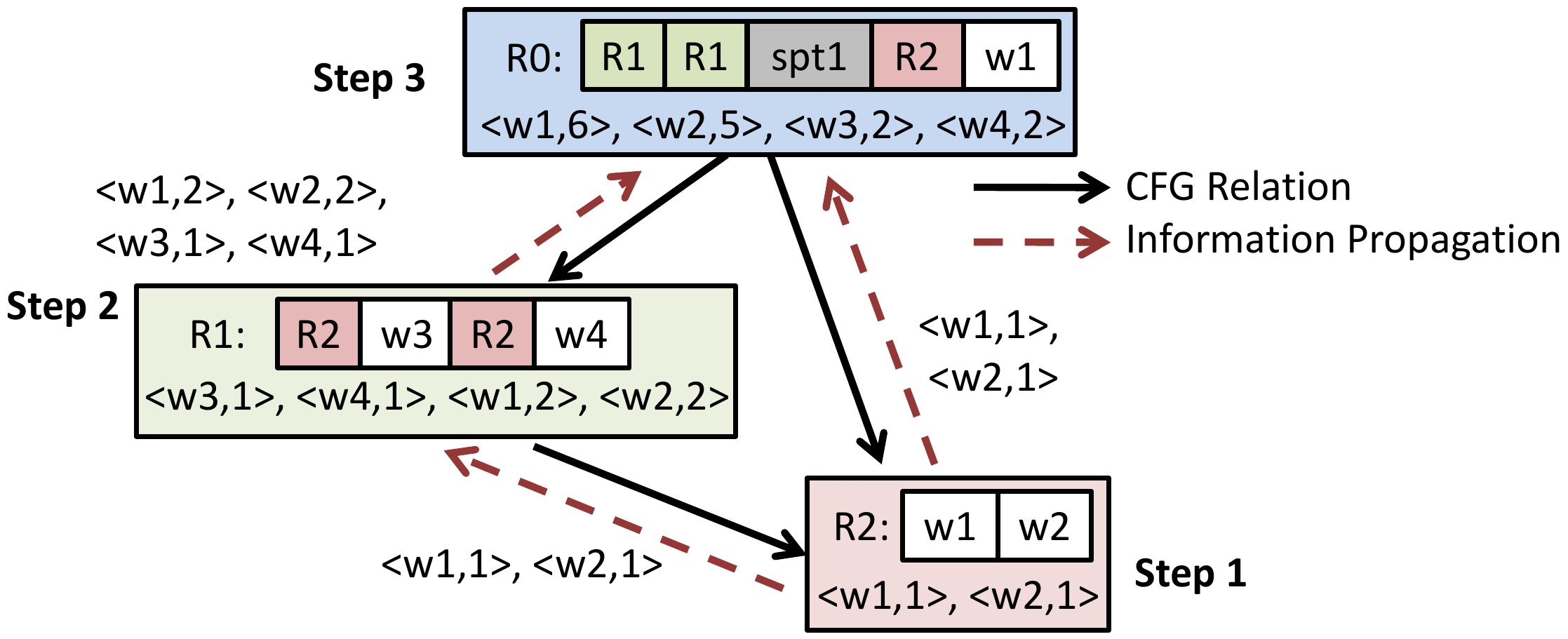}

  \caption{Word count example using TADOC.}
  \label{fig:wordcount}

\end{figure}

We use Figure~\ref{fig:wordcount} to show how to utilize the TADOC compressed data to perform a simple data analytics application -- \emph{word count}.
In Step 1,
\emph{R2} transmits its accumulated local word frequencies to its parents,
which are \emph{R0} and \emph{R1}.
In Step 2,
\emph{R1} receives the word frequencies from \emph{R2} and merges these frequencies to \emph{R1}'s local frequency table.
In Step 3,
\emph{R1} transmits its accumulated word frequencies to its parent \emph{R0}.
After \emph{R0} receives the word frequencies from all its children,
which are \emph{R1} and \emph{R2},
\emph{R0} merges all received word frequencies into \emph{R0}'s word count results, which are also the final word counts.

\subsection{GPU}

GPU is a specialized device targeting graphics and image processing originally.
Due to its high parallelism design,
GPUs have now been applied to a wide range of applications,
including data management systems~\cite{root2016mapd,yuan2016spark,koliousis2016saber,wang2017gunrock,zhang2020finestream}.
A server equipped with GPUs can offer unprecedented computing power within a single machine.
Previous TADOC mainly targets distributed environments without GPUs.
If we provide a GPU solution for TADOC, enabling efficient GPU-based data analytics without decompression,
not only the number of machines, but also the electricity \cb{budget} can all be saved.

GPUs are different from CPUs in various aspects.
First, different from CPUs,
GPUs include a large number of light-weight cores,
grouped into different streaming multi-processors (SM).
GPUs utilize high throughput to hide memory access latency.
Second,
each SM has controllable shared memory,
which has similar performance to caches.
Therefore,
good use of shared memory is critical for GPU performance.
Third,
the execution model on GPUs is also quite different from that on CPUs.
The basic thread scheduling unit is a \emph{warp}, which includes 32 threads for Nvidia GPUs.
The threads within the same warp execute the same instructions in a lock-step manner, so called single instruction, multiple threads (SIMT).
Developing GPU-based TADOC needs to adapt to the GPU execution model.

\section{Motivation}

\subsection{Revisiting Previous Techniques}
In this part, we revisit previous traversal-based techniques to show our motivation of a new GPU-based TADOC design.
\vspace{\vsp{}}

\cb{
\textbf{Why GPU-based BFS does not apply?}
The DAG traversal in TADOC is unique and cannot be replaced by a BFS traversal.
First,
many data analytics tasks require sequence maintenance of words, where BFS cannot be used directly~\cite{zhang2018efficient}.
Second,
TADOC involves complicated data processing and complex data structures during traversal.
For example,
each rule, which is a  node in DAG, needs to maintain a local word table and a rule table, and all rules write to the same global buffer, which generates write conflicts in G-TADOC among GPU threads.
Third,
the DAG traversal in TADOC involves dynamic data transmission.
For example, the traversal can transmit accumulated word frequencies among rules.
Unfortunately, the amount of data transferred between nodes cannot be obtained in advance,
which has not been involved in BFS on GPUs.
}
\vspace{\vsp{}}

\cb{
\textbf{Why existing DAG traversal on GPUs
 does not apply?}
The uniqueness of TADOC is that each node requires  complicated text-related intra- and inter-node operations.
This uniqueness does not need to be considered in previous GPU traversal solutions.
In detail, within a node, a dynamic buffer needs to be maintained to receive intermediate results from  parents and to transmit data to children.
Between different nodes, 
cross-rule sequence needs to be considered.
}

\subsection{Challenges}
\label{sec:challenge}

In this section,
we mainly discuss the challenges of enabling efficient text analytics on GPUs without decompression.

\vspace{\vsp{}}
\textbf{Challenge 1: GPU parallelism for TADOC}.
The high performance of GPU relies on the high throughput from thread-level parallelism.
First, as presented in~\cite{zhang2018efficient},
there exist massive dependencies among the DAG,
which leads TADOC difficult to be parallel.
Accordingly, TADOC utilizes coarse-grained parallelism that mainly processes different compressed files in parallel: each CPU thread handles a separate file~\cite{zhang2020tadoc}.
We cannot apply such coarse-grained parallelism on GPUs because a GPU supports thousands of threads and it is inefficient to split the compressed data into that large number of partitions.
Second,
if we use one GPU thread for one rule,
there is a workload unbalancing problem because the numbers of elements in different rules vary significantly.
GPUs launch threads in warp level, and the threads within a warp have to release resources simultaneously.
The workload imbalance problem decreases the parallelism degrees.
Third,
we cannot simply decide the number of threads for rules,
because of the various rule length.

\vspace{\vsp{}}


\textbf{Challenge 2: TADOC final result update conflict of massive GPU threads}.
The update conflict is a serious problem when we develop TADOC on GPUs.
First,
the update conflict of multiple threads writing to the same result buffer is not a serious problem on CPUs because the number of CPU threads is limited.
However,
on a GPU server,
when a large number of GPU threads write to the same result buffer,
we have to use atomic operations to guarantee correctness,
which incurs massive conflicts.
Second,
the complicated data structures used in TADOC cannot be applied in GPU environment.
For example, TADOC uses an \emph{unordered map} data structure for results such as word counts;
we need to develop our own similar data structures on GPUs with atomicity and consistency considered.
Third,
the amount of memory required by TADOC 
is unknown until runtime.
Even worse,
for TADOC on GPUs,
the memory sizes of different threads are also various,
which makes the update problem with thread conflicts more difficult.

\vspace{\vsp{}}

\textbf{Challenge 3: sequence maintenance of TADOC compressed data on GPUs}.
How to keep the sequence information on GPUs is also challenging.
Sequence maintenance is essential for sequence sensitive applications,
such as counting three continuous word sequences. 
First,
to keep the sequence information,
TADOC originally traverses the DAG in a DFS order~\cite{zhang2018efficient},
which is hard to be parallel.
Second,
a word sequence can span several rules and these rules can be controlled by different GPU threads.
Currently,
threads across different GPU blocks have no mechanism for synchronization.
Third,
TADOC uses \emph{map} data structures to store sequence counts.
For these sequence-based applications,
we need to develop special data structures in GPU memories to store sequences and  perform basic comparisons between threads.

Based on the analysis, designing a GPU-based TADOC is very rewarding, but full of challenges.

\section{\sysname{}}
In this section, we show our \sysname{} framework.
\sysname{} consists of three components,
a module for data structures,
a parallel execution module,
and a sequence support module.
We first show the \sysname{} overview and then the different modules.

\subsection{Overview}

\begin{figure}[h!]

\includegraphics[width=1\linewidth]{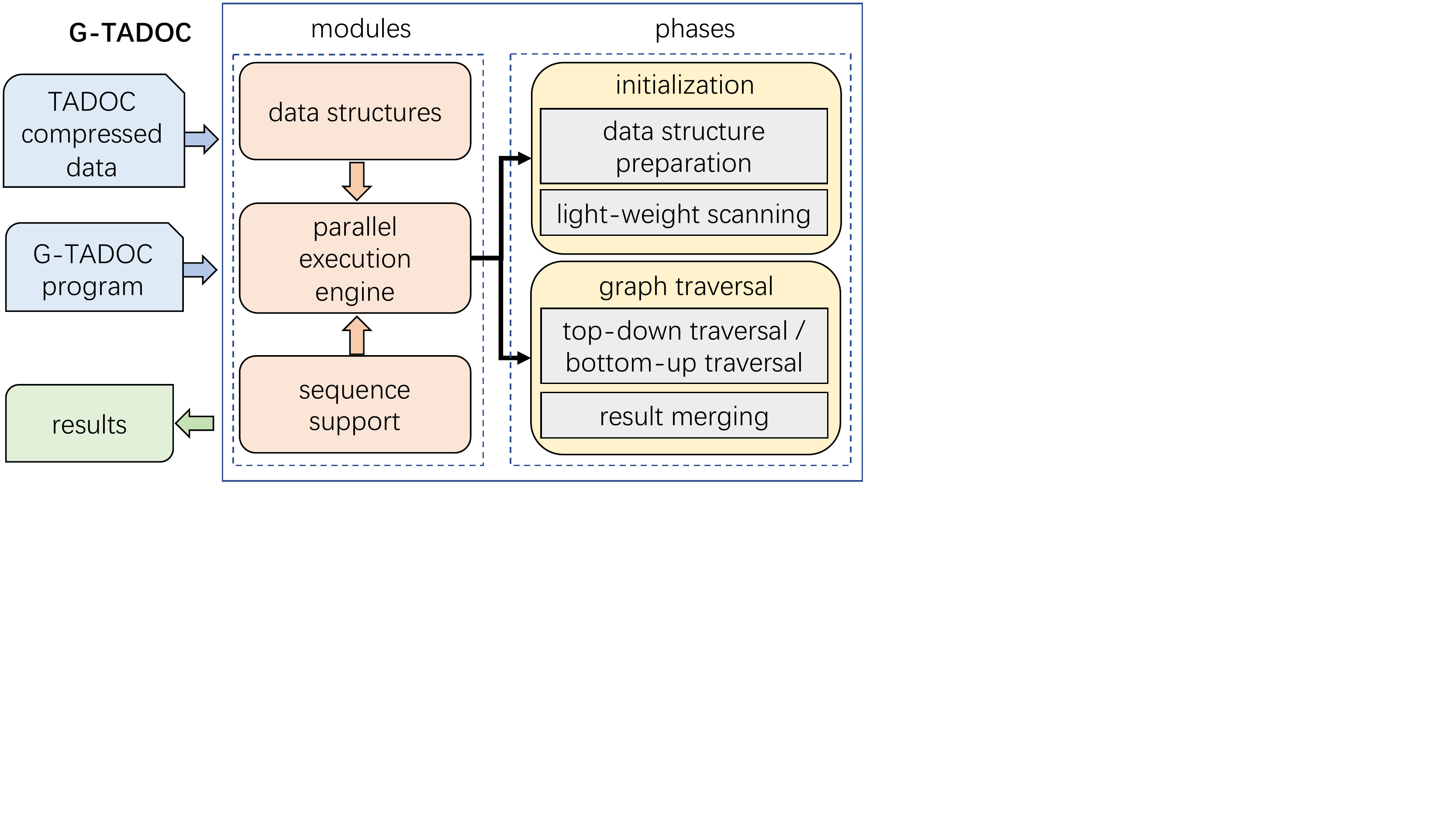}

  \caption{\sysname{} overview.}
  \label{fig:ovev}

\end{figure}

We show the  overview of \sysname{} in Figure~\ref{fig:ovev}.
The inputs are TADOC compressed data and user program, and the outputs are the results,
which are similar to previous non-GPU TADOC implementations~\cite{zhang2018efficient,Zhang2018ics}.

\vspace{\vsp{}}

\textbf{Modules}.
\sysname{} consists of three major modules.
The parallel execution module is responsible for the \sysname{} parallel execution on GPUs,
which decides how to partition workloads for thread parallelism.
The data structure module provides necessary data structures for \sysname{} execution,
including a self-controlled memory pool,
thread-safe data structures,
and \emph{head} and \emph{tail} structures for sequences.
The sequence support module is used for applications that are sensitive to sequence orders.

\vspace{\vsp{}}

\textbf{Phases}.
After receiving the TADOC compressed data and program,
\sysname{} execution mainly consists of two phases: initialization phase and graph traversal phase.
In the initialization phase,
\sysname{} prepares necessary data structures according to the user program and launches a light-weight scanning to fulfill related values.
In the graph traversal phase,
\sysname{} analyzes different traversal strategies and chooses the most suitable one based on both data and tasks.
Before the end of the graph traversal, \sysname{} performs a merging process for final results.

\vspace{\vsp{}}

\textbf{Solutions to challenges}.
\sysname{} can handle the challenges mentioned in Section~\ref{sec:challenge}.
To address the first GPU parallelism challenge,
\sysname{} adopts a thread-level workload scheduling strategy for GPU threads, which partitions the DAG in a fine-grained manner for parallelism
(Section~\ref{subsec:paraExe}).
To address the second TADOC update conflict challenge,
we develop a memory pool on GPUs and maintain necessary data structures so that all threads manage the same memory objects with consistency guaranteed (Section~\ref{subsec:dataStructure}).
To address the third challenge of sequence sensitivities on GPUs,
\sysname{} scans the DAG for recording the cross-rule content in a light-weight manner in the initialization phase.
Then, \sysname{} performs a rule-level processing and result merging process in the graph traversal phase (Section~\ref{subsec:seqSupport}).

\subsection{Fine-Grained Thread-Level Execution Engine}
\label{subsec:paraExe}

\begin{figure}[h!]

\centering
\includegraphics[width=0.85\linewidth]{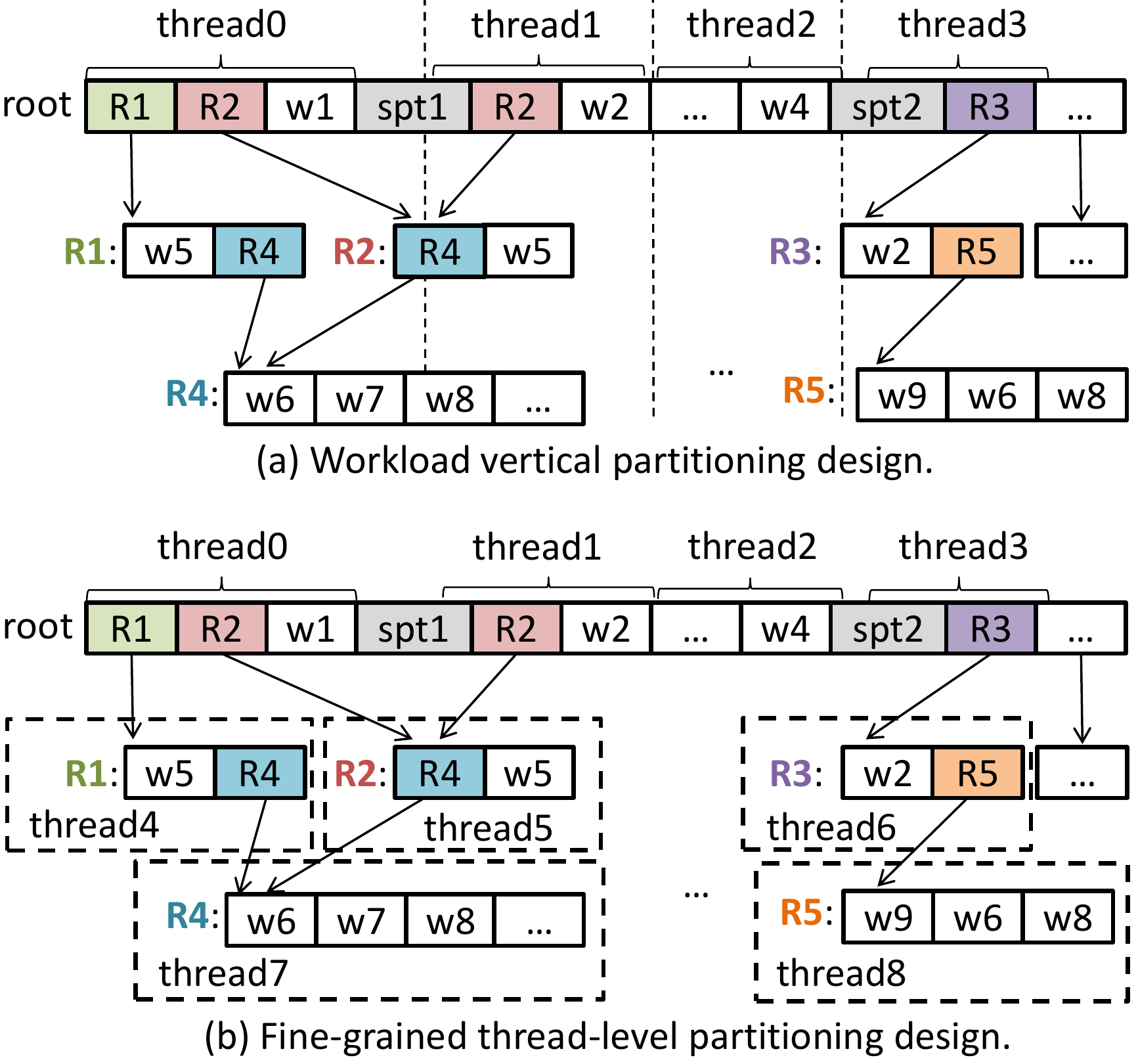}

  \caption{Workload partitioning design exploration.}
  \label{fig:indexing}

\end{figure}

We show our \sysname{} parallel execution engine in this part.
In developing our parallel partitioning strategy,
we consider two possible designs, as shown in Figure~\ref{fig:indexing}.
The first design is to partition the DAG vertically from the root: different parts are traversed by different threads,
as shown in Figure~\ref{fig:indexing}~(a).
This design can leverage the GPU parallelism,
but at the same time, some rules can be scanned by different threads.
For example, \emph{R2} and \emph{R4} are scanned by both \emph{thread0} and \emph{thread1}.
Even worse, when the DAG is very deep and complicated,
the problem that massive rules are repeatedly scanned by different threads can be serious.
Hence, we abandon this design.
The second design is fine-grained thread-level scheduling:
we assign a thread for each node except the root;
the root rule usually includes a large number of elements so we allocate a group of threads based on the rule length to handle it.
Note that when a rule includes a large number of elements (the default threshold is 16 times the average number of elements per thread), such as \emph{R4},
more threads should be allocated for the rule.
To traverse the DAG,
each rule is associated with a \emph{mask} to indicate whether a rule is ready to be traversed or not.
This design  ensures the dependency for correctness in the DAG traversal and retains great parallelism simultaneously.
Therefore, we adopt this fine-grained design.
Moreover,
as discussed in~\cite{zhang2020tadoc},
the optimal traversal strategy depends on both input data and analytics tasks,
so we develop both top-down and bottom-up traversals and use the strategy selector in~\cite{zhang2020tadoc} for such decisions.

Next, we show our detailed top-down and bottom-up designs in \sysname{}.

\vspace{\vsp{}}

\vspace{0.05in}

\textbf{Top-down traversal}. 
\cb{
We show our top-down traversal design in this part.
}

\cb{
\emph{1) General design}.
The general design of top-down traversal transmits required data, such as file information, from the root to sub-nodes for processing. Then, \sysname{} gathers local results from different nodes as the final result.
First,
in root,  different consecutive parts are controlled by different threads, which can be processed in parallel.
Second,
during traversal, multiple parents can write to the same local buffer of a rule,
which relates to data consistency.
To handle this problem, a self-maintained memory pool is introduced, detailed in Section~\ref{subsec:dataStructure}.
Third,
G-TADOC reduces intermediate results from rules to the final output in a global buffer in parallel.
}

\emph{\cb{2) Detailed algorithm}}.
We show our top-down DAG traversal design in Algorithm~\ref{alg:topdown}.
The general control is performed on the CPU side by the \emph{topDown} function,
as shown in Lines~\ref{topdown:start1} to~\ref{topdown:end1},
which calls different GPU kernels, including \emph{initTopDownMaskKernel}, \emph{topDownKernel}, and \emph{reduceResultKernel}.

{
The \emph{initTopDownMaskKernel} is executed on GPUs, which initializes the nodes whose in-edges are from only the root. 
In detail, we  consider the topology of the rules except the root, and accordingly, the initial weights of these rules are their frequencies in the root.
Additionally, \emph{rule.numInEdge} stores the number of in-edges of each rule, and only the rules with zero \emph{numInEdge} can start the DAG traversal initially.
In \sysname{},
we mark the masks of the rules that can be processed as \emph{true}.
}

The \emph{topDownKernel} is the main body of the top-down traversal and is executed on GPUs.
We set the \emph{devStopFlag} to \emph{true} in Line~\ref{topdown:stopflag1}.
If the \emph{devStopFlag} is still \emph{true}  after the \emph{topDownKernel} execution,
 which means that the DAG has no update and has been fully traversed,
 then \sysname{} stops traversal.
 In \emph{topDownKernel},
 for different applications, only the for-loop from Lines~\ref{topdown:innerfor1} to~\ref{topdown:atomicadd} 
 is different. Here, we take \emph{word count} as an example.
 For a given rule,
 it first transmits its accumulated weights to all its subrules (Line~\ref{topdown:atomicadd}).
 If the number of current in-edges \emph{subRule.curInEdge} is equal to a subrule's full number of in-edges \emph{subRule.numInEdge},
 then we mark the subrule's mask to \emph{true},
 indicating that the subrule is ready to be traversed in the next round (Line~\ref{topdown:masktrue}).
 Note that when any masks are changed, \emph{stopFlag} shall be set to \emph{false}.
 Moreover, \emph{rule.mask} should be set to \emph{false} in Line~\ref{topdown:maskfalse} so that the rule will not be involved in the next round.

The \emph{reduceResultKernel} merges the word frequencies from all rules multiplied by their corresponding accumulated rule weights on GPUs.

\algdef{SE}[DOWHILE]{Do}{doWhile}{\algorithmicdo}[1]{\algorithmicwhile\ #1}%

\begin{algorithm}[!ht]
  \caption{Top-Down Traversal}
  \label{alg:topdown}
  \footnotesize
  \begin{algorithmic}[1]
    \algrenewcommand\textproc{}
    \Function{topDown}{$rules$} \Comment{Executed by host; use \emph{word count} as an example}
    {\label{topdown:start1}}
      \State{initTopDownMaskKernel$(rules)$ with at least $rules.size$ threads}
            {\label{topdown:initmask}}
      \Do \Comment{Repeated top-down traverse until all rules' weight generated} 
            {\label{topdown:dowhilestart}}
        \State{cudaMemSet $devStopFlag \leftarrow true$}
            {\label{topdown:stopflag1}}
        \State{topDownKernel$(rules, devStopFlag)$ with at least $rules.size$ threads}
        \State{cudaMemCpy $devStopFlag$ to $stopFlag$}
      \doWhile{$stopFlag$ is $false$}
            {\label{topdown:dowhileend}}
      
        \State{reduceResultKernel$(rules)$ with at least $rules.size$ threads} \Comment{Reduce results from all rules}
      
          {\label{topdown:end1}}
        
    \EndFunction

    \Statex{}
    
    \Function{topDownKernel}{$rules, devStopFlag$} \Comment{GPU kernel} 
      \If {$tid$ not in $1$ to $rules.size-1$}
        \State {\Return}
      \EndIf
      \State $rule \leftarrow rules[tid]$
      \If {$rule.mask$ is $false$}
        \State {\Return} 
      \EndIf
      \For {\texttt{each $subRuleId, subRuleFreq$ in $rule.subRules$}} 
                  {\label{topdown:innerfor1}}
        \State {$subRule \leftarrow rules[subRuleId]$}
        \State {atomicAdd$(subRule.weight, subRuleFreq * rule.weight)$} 
                          {\label{topdown:atomicadd}}
        \State {atomicAdd$(subRule.curInEdge, 1)$} 
        \If {$subRule.curInEdge$ is full}
          \State {$subRule.mask \leftarrow true$} \Comment{Sub-rule then can be traversed}
                            {\label{topdown:masktrue}}
          \State {$devStopFlag \leftarrow false$}
                                      {\label{topdown:stopfalse}}
        \EndIf
      \EndFor
                  {\label{topdown:innerfor2}}
      \State{$rule.mask \leftarrow false$}
            {\label{topdown:maskfalse}}
    \EndFunction

  \end{algorithmic}
\end{algorithm}

\vspace{\vsp{}}

\cb{
\emph{3) Complexity analysis}.
Algorithm~\ref{alg:topdown} can be divided into three stages.
The first stage is mask initialization (Line~\ref{topdown:initmask}),
in which each thread checks the corresponding rule's number of in-edges and then sets its mask.
Assuming sufficient parallel resources,
the complexity is $O(1)$.
The second stage is top-down traversal (Lines~\ref{topdown:dowhilestart} to~\ref{topdown:dowhileend}).
 Assuming that the DAG has $k$ layers,  the number of loops is not greater than $k$. Then each thread in \emph{topDownKernel}  traverses the corresponding rule's sub-rules. Suppose in the $i^{th}$ loop, the maximum number of sub-rules of a rule  is $e_{i, max}$, then the total complexity of this stage is $O(\sum_{i=1}^{k}e_{i, max})$, which can be represented as $O(k\Bar{e}_{max})$.
 The third stage is to reduce results (Line~\ref{topdown:end1}). Each thread needs to merge the corresponding rule's local words from the local table to the global table, so the complexity is $O(w_{max})$, where $w_{max}$ is the maximum number of local words among all rules.
 Therefore, the overall complexity of Algorithm~\ref{alg:topdown} is $O(k\Bar{e}_{max}+w_{max})$.
}

\vspace{0.05in}

\textbf{Bottom-up traversal}.
\cb{
We show our bottom-up traversal design in this part.
}

\cb{
\emph{1) General design}.
The bottom-up traversal transmits required data, such as local word counts, from leaves to upper-level nodes.
After transmission,
the root and its directly connected nodes (called 2nd-layer nodes) store the gathered result.
Note that we do not accumulate the results to the root because the root contains file information.
In detail,
first, each leaf transmits the required data from its local tables to its parents.
Second, during traversal,
each node accumulates the transmitted data from children and then transmits the accumulated results to its parents.
Note that data consistency needs to be guaranteed since different rules are controlled by different threads.
Third,
after traversal, G-TADOC analyzes the local buffers in the root and 2nd-layer nodes in parallel to generate the final results.
}

\cb{\emph{2) Detailed algorithm}}.
We show our bottom-up DAG traversal design in Algorithm~\ref{alg:bottomup}.
The general control is performed by the function \emph{bottomUp} from the CPU side, 
which calls different GPU kernels.
Different from Algorithm~\ref{alg:topdown},
the bottom-up design in Algorithm~\ref{alg:bottomup} first generates the pointers from children to parents (Lines~\ref{bottomup:allocmem1} to~\ref{bottomup:ctop}),
initializes masks (Line~\ref{bottomup:maskinit}),
and generates the local tables' bound in a light-weight bottom-up manner (Lines~\ref{bottomup:do} to~\ref{bottomup:endpan}), so that the local tables in rules can be allocated (Line~\ref{bottomup:allocmem2}).
Then,
it initializes masks again (Line~\ref{bottomup:maskinit2})
and traverses the graph in a comprehensive bottom-up direction with a result merging process (Lines~\ref{bottomup:dopan} to~\ref{bottomup:end1}).

\cb{
The \emph{initBottomUpMaskKernel} set the rule masks. The leaves are set to \emph{true} so that they can be traversed initially. 
}

The \emph{genLocTblBoundKernel} is used to calculate the memory size limit for local tables, and is called by the \emph{bottomUp} function repeatedly.
\cb{
Its kernel execution is similar to that of \emph{topDownKernel} in Algorithm~\ref{alg:topdown}, except the use of out-edge rather than in-edge during traversal.
When a rule is traversed, G-TADOC sums the upper limits of its local words and all its children's local tables as the amount of space that should be allocated.
Then, the rule  increases all its parents' out-edges. When a parent's number of current out-edges is equal to its number of subrules, G-TADOC sets its mask to \emph{true} for the next-iteration execution. 
After calculating the memory limit of each node, we uniformly allocate the corresponding buffer for each rule in $rules.locTbl$ (Line~\ref{bottomup:allocmem2}).
}

The \emph{genLocTblKernel} is used for DAG traversal with the allocated memory space from the \emph{bottomUp} function.
\cb{
Its traversal order is controlled by the traversed out-edges,
which is the same as \emph{genLocTblBoundKernel}.
However, the kernel's computation task is much heavier.
Here, we use the \emph{word count} example for illustration.
When a rule is traversed, it  first reduces its local word frequencies, and then merges all its subrules' local word frequencies into its own local table.
}

 The \emph{reduceResultKernel} merges the word frequencies from the root and its children where the root is directly connected (called \emph{level-2 nodes} in~\cite{zhang2018efficient}) on GPUs.
In detail, \sysname{} merges 1) the word frequencies in the root, and 2) the frequencies in the local tables of the root's direct children multiplied by their corresponding rule frequencies in the root. \cb{This is different from the \emph{reduceResultKernel} in Algorithm~\ref{alg:topdown}.
}

\begin{algorithm}[!h]
  \caption{Bottom-Up Traversal}
  \label{alg:bottomup}
  \footnotesize
  \begin{algorithmic}[1]
    \algrenewcommand\textproc{}
    
        \Function{bottomUp}{$rules$} \Comment{\emph{word count}}
                  {\label{bottomup:start1}}

      \State{allocate device memory to $rules.parentIds$}
                        {\label{bottomup:allocmem1}}
      \State{genRuleParentsKernel$(rules)$ with at least $rules.size$ threads}
                        {\label{bottomup:ctop}}
      \Statex{}
      \State{initBottomUpMaskKernel$(rules)$ with at least $rules.size$ threads}
                              {\label{bottomup:maskinit}}

      \Do 
                                    {\label{bottomup:do}}

        \State{cudaMemSet $devStopFlag \leftarrow true$}
        \State{genLocTblBoundKernel$(rules, devStopFlag)$ with at least $rules.size$ threads}
        \State{cudaMemCpy $devStopFlag$ to $stopFlag$}
      \doWhile{$stopFlag$ is $false$}
                                {\label{bottomup:endpan}}
      
      \State{allocate device memory to $rules.locTbl$}
                                {\label{bottomup:allocmem2}}
      
      \Statex{}
      \State{initBottomUpMaskKernel$(rules)$ with at least $rules.size$ threads}
                                {\label{bottomup:maskinit2}}
      \Do 
                                    {\label{bottomup:dopan}}
        \State{cudaMemSet $devStopFlag \leftarrow true$}
        \State{genLocTblKernel$(rules, devStopFlag)$ with at least $rules.size$ threads}
        \State{cudaMemCpy $devStopFlag$ to $stopFlag$}
      \doWhile{$stopFlag$ is $false$}
      
      \State{reduceResultKernel$(rules)$ with at least $root.size$ threads}

                        {\label{bottomup:end1}}
    \EndFunction

  \end{algorithmic}
\end{algorithm}

\cb{
\emph{3) Complexity analysis}.
Different from Algorithm~\ref{alg:topdown},
Algorithm~\ref{alg:bottomup} consists of five stages.
The first stage is to generate the parents of rules (Lines~\ref{bottomup:allocmem1} to~\ref{bottomup:ctop}). Each thread in \emph{genRuleParentsKernel}  stores the corresponding rule's ID in all its sub-rules' parent table. The complexity is $O(e_{max})$, where $e_{max}$ is the maximum number of sub-rules of all rules.
The second stage is mask initialization (Line~\ref{bottomup:maskinit} and~\ref{bottomup:maskinit2}). Similar with the mask initialization in Algorithm~\ref{alg:topdown},  the complexity is also $O(1)$.
The third stage is  to generate rules' local table bound (Lines~\ref{bottomup:do} to~\ref{bottomup:endpan}). 
Each thread in \emph{genLocTblBoundKernel}  traverses the corresponding rule's sub-rules and parents. Suppose in the $i^{th}$ loop, the maximum numbers of sub-rules and parents of these rules are $e_{i, max}$ and $p_{i, max}$ respectively. Then, the complexity is $O(\sum_{i=1}^{k}(e_{i, max}+p_{i, max})) = O(k(\Bar{e}_{max}+\Bar{p}_{max}))$, where $k$ is the number of layers in the DAG.
The fourth stage is to generate rules' local table (Lines~\ref{bottomup:do} to~\ref{bottomup:endpan}). 
Besides traversing corresponding rule's sub-rules and parents, each thread in \emph{genLocTblKernel}  also merges all sub-rules' local tables and its own words. 
For a given rule $i$, suppose its local table size is $t_i$ and its number of words  is $w_{i}$, then its computation load is $w_i+\sum_{j\in i.subRules}t_j$. 
The complexity of this stage is $O(\sum_{i=1}^{k}C_{i, max})$, which is $O(k\Bar{C}_{max})$. $C_{i, max}$ is the maximum computation load among rules in the  $i^{th}$ loop.
The fifth stage is to reduce results (Line~\ref{topdown:end1}). This stage scans the root and merges all \emph{level-2 nodes}. In detail, each thread is responsible for one \emph{level-2 node}, so the complexity is $O(t_{lv2, max})$, where $t_{lv2, max}$ is the maximum size of \emph{level-2 nodes'} local tables.
Therefore,
the overall complexity of Algorithm~\ref{alg:bottomup} is $O(k(\Bar{e}_{max}+\Bar{p}_{max}+\Bar{C}_{max})+t_{lv2, max})$.
}

\vspace{0.1in}

\vspace{0.05in}

\cb{
\textbf{Parameter selection}.
G-TADOC involves a few parameters to adjust,
such as the threshold of GPU thread resources allocated to a rule.
The current solution is to extract a sample set of input and then use a greedy strategy to set each parameter in turns.
If the input is unavailable until runtime,
then the parameters are set according to our training set (a small extracted dataset from Wikipedia~\cite{wikipedia}).
}

\subsection{\sysname{} Data Structures}
\label{subsec:dataStructure}

The data structures in \sysname{} include a self-maintained memory pool, thread-safe structures,
and sequence support.

\vspace{\vsp{}}

\textbf{\sysname{} maintained memory pool}.
As discussed in Section~\ref{subsec:paraExe},
we need to provide each thread a separate memory space during DAG traversal.
Because 1) the required memory size is unknown until runtime, and 2) allocating memory dynamically for all threads is inefficient,
we develop a global memory pool to manage the GPU memory by \sysname{} itself.
First,
each rule calculates its own required memory size for necessary data structures.
Second,
with data transmission in the initialization phase in Figure~\ref{fig:SequiturExample},
each rule transmits its memory requirement to its parents in a bottom-up traversal, or to its children in a top-down traversal. This memory requirement transmission process can be recursive.
Third,
after the whole range transmission in the initialization phase,
each rule determines its maximum memory requirement and we can allocate related resources of different rules from the memory pool.

\vspace{\vsp{}}

\begin{figure}[h!]

\centering
\includegraphics[width=0.95\linewidth]{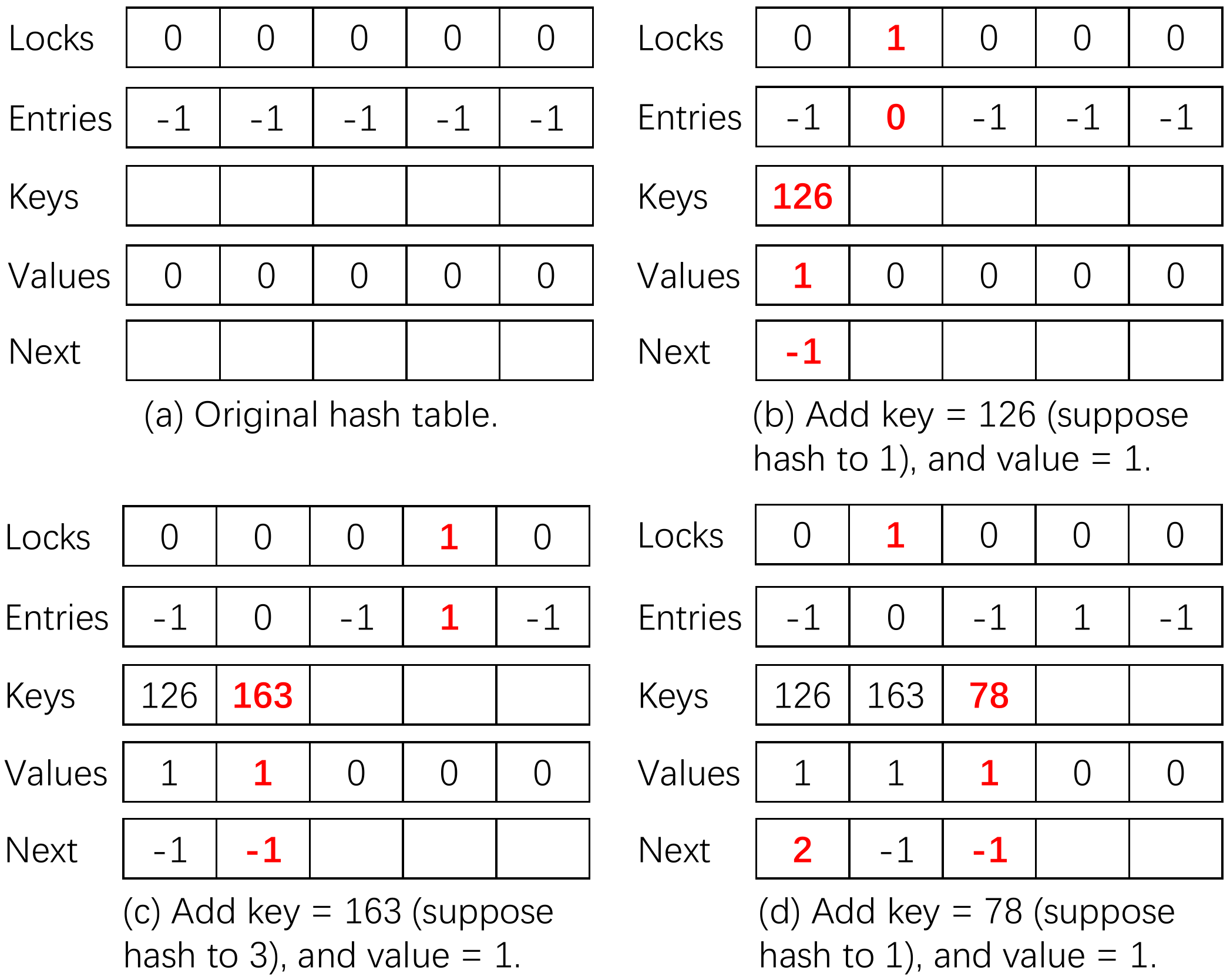}

  \caption{Illustration for thread-safe hash tables.}
  \label{fig:hash}

\end{figure}

\textbf{Thread-safe data structures}.
After we introduce the memory pool in \sysname{},
we next describe the thread-safe data structures used in the memory pool for GPU threads.
The most important data structure in TADOC is the \emph{hash} structure~\cite{zhang2018efficient},
which can be used to store the results both locally and globally.
Hence, 
we use the hash structure for illustration in \sysname{} thread-safe design, as shown in Figure~\ref{fig:hash}.
The original state of the hash table is shown in Figure~\ref{fig:hash}~(a).
The lock buffer is for locking entries (1 means locked, and 0 means unlocked).
The entry buffer is for hashing (default -1).
The key and value buffers are for the key-value pairs.
The \emph{next} buffer is for the next entry if multiple key-value pairs are mapped into the same entry.
Figure~\ref{fig:hash}~(b) shows the state after inserting \textless126,1\textgreater, assuming the key-value pair is hashed to 1.
Because there is no conflict in this insertion,
the related value in the \emph{next} buffer is -1.
Figure~\ref{fig:hash}~(c) shows the hash table state after inserting  \textless163,1\textgreater, assuming the key-value pair is hashed to 3.
Accordingly, \sysname{} just stores the key-value pair $\textless163,1\textgreater$  after the first \textless126,1\textgreater.
Figure~\ref{fig:hash}~(d) shows a hash conflict situation: the hash table state after inserting  \textless78,1\textgreater, assuming the key-value pair is hashed to 1.
Because $\textless126,1\textgreater$  has already been inserted to the first entry,
we update its ``next'' buffer pointing to a new place for the newly inserted \textless78,1\textgreater.
Note that the lock buffer is used only when all threads writing to the same buffer location.
Moreover, if the hash table is private and owned by one thread,
we do not need to create the locks.

\vspace{\vsp{}}

\begin{figure}[h!]

\centering
\includegraphics[width=0.75\linewidth]{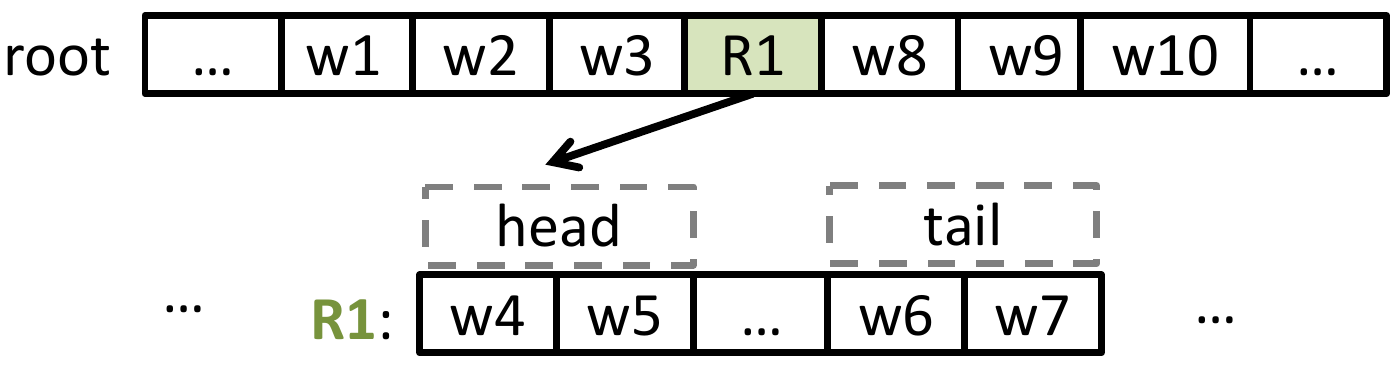}

  \caption{Head and tail data structures for sequences.}
  \label{fig:headtail}

\end{figure}

\textbf{Head and tail structures for sequence support}.
The head and tail structures are used to support sequence sensitive applications, such as \emph{sequence count}~\cite{zhang2018efficient}.
Because \sysname{} traverses the DAG in parallel,
some rules may involve cross-rule sequence (a word sequence spanning multiple nodes in the DAG).
We design \emph{head} and \emph{tail} data structures for each rule to store the content of the beginning and end of the rule,
which are provided to the parents.
We show an example in Figure~\ref{fig:headtail}.
In the root, the first sequence, {\textless}\emph{w1,w2,w3}\textgreater, is a  sequence that does not span across rules.
However, for the next three-word sequence, {\textless}\emph{w2,w3,w4}\textgreater,
it spans across the root and \emph{R1}.
For this sequence, we store the partial content of ${\textless}\emph{w4,w5}\textgreater$  in the head buffer of \emph{R1},
so that this cross-rule sequence can be processed by the parent, which is the root.
Similarly,
we store ${\textless}\emph{w6,w7}\textgreater$ in the tail buffer of \emph{R1} so that \emph{R1}'s parent can quickly process the sequences containing the words in \emph{R1}'s tail buffer.
Note that  the first few elements and the last few elements in the subrule can also be a rule.
For example, 
in Figure~\ref{fig:indexing}, the first element of \emph{R2} is also a rule,
so the sequence from the root can span more than two rules,
which is complicated.
In our design, each rule can be handled by different threads.
If we can provide the head and tail buffers of all rules,
we can avoid multi-rule scanning by looking into only the head and tail buffers  of different subrules directly.
In summary,
the parents are responsible to process cross-rule sequences,
and the problem can be solved by scanning the head and tail buffers of the direct children.
More details are presented in Section~\ref{subsec:seqSupport}.

\subsection{Sequence Support in \sysname{}}
\label{subsec:seqSupport}

In this part, we discuss the sequence support in \sysname{} for sequence sensitive applications.
The sequence support in TADOC~\cite{zhang2018efficient} is developed by function recursive calls,
which is inefficient and hard to be parallel on GPUs.
To improve the sequence support of TADOC and parallelize it on GPUs,
we have the following insights.
First,
to fully parallelize the rule processing,
each rule needs to include the head and tail buffers mentioned in
Section~\ref{subsec:paraExe} to remove the sequence dependency across rules.
Second,
a first-round initialization phase is required to fulfill the head and tail buffers for all rules.
Third,
the original recursive design in TADOC~\cite{zhang2018efficient} is inefficient and thus shall be abandoned; a more efficient parallel graph traversal needs to be developed.

Based on the analysis,
we develop a two-phase sequence support design for sequence sensitive applications.

\vspace{\vsp{}}

\textbf{Initialization phase}.
The first initialization phase is to prepare the head and tail buffers for each rule with a light-weight scanning.
The upper limit of memory space for each rule is shown in Equation~\ref{eq:upperbound},
where \emph{wordSize} denotes the size of the word elements,
\emph{l} denotes the sequence length, and
\emph{subRuleSize} denotes the number of subrules.

  \begin{equation}
\resizebox{1\linewidth}{!}{
    \label{eq:upperbound}
    {$
    upperLimit=wordSize+(l-1)\times subRuleSize-(l-1)
$}
  }
  \end{equation}

\begin{figure}[!h]

\includegraphics[width=1\linewidth]{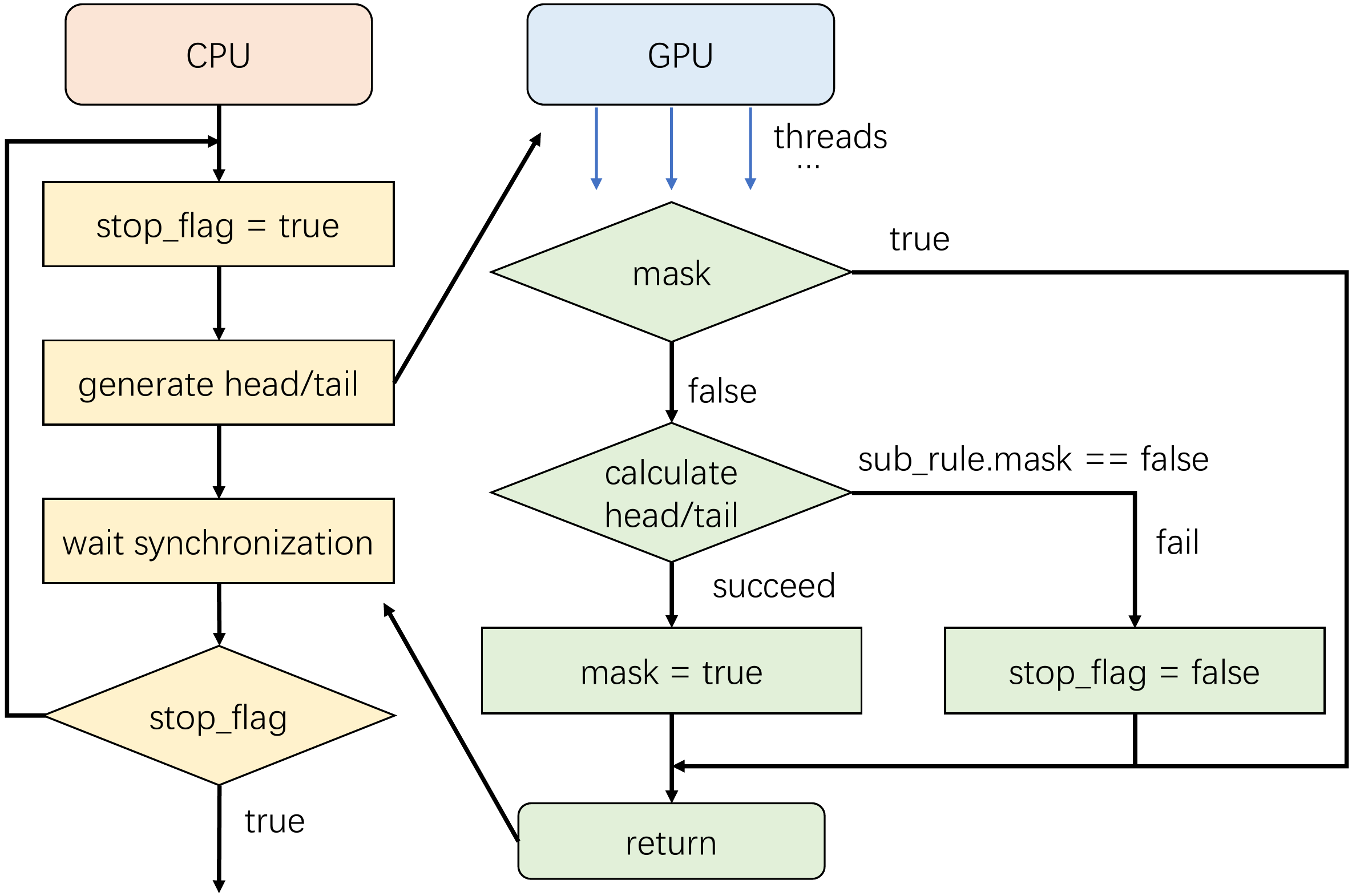}

  \caption{Phase 1: initialization for head and tail buffers.}
  \label{fig:scan1}

\end{figure}

The detailed process to generate the head and tail buffers of each rule is shown in Figure~\ref{fig:scan1}.
The CPU side uses a while-loop to continuously check whether all the head and tail buffers have been fulfilled.
To generate the head buffers,
\sysname{} traverses the rules, and puts a given number of continuous words at the beginning of the rule in the head buffer.
Within such a process,
if \sysname{} encounters a subrule,
\sysname{} first checks the related mask.
If the mask is set,
which implies that the subrule's head buffer is ready,
then \sysname{} can put the content from the subrule's head buffer to the current rule's head buffer;
otherwise,
 the calculation fails and needs 
 to be conducted in the
 next round.
The generation of the tail buffers is similar to the generation of the head buffers.

\vspace{\vsp{}}

\begin{figure}[!h]

\includegraphics[width=1\linewidth]{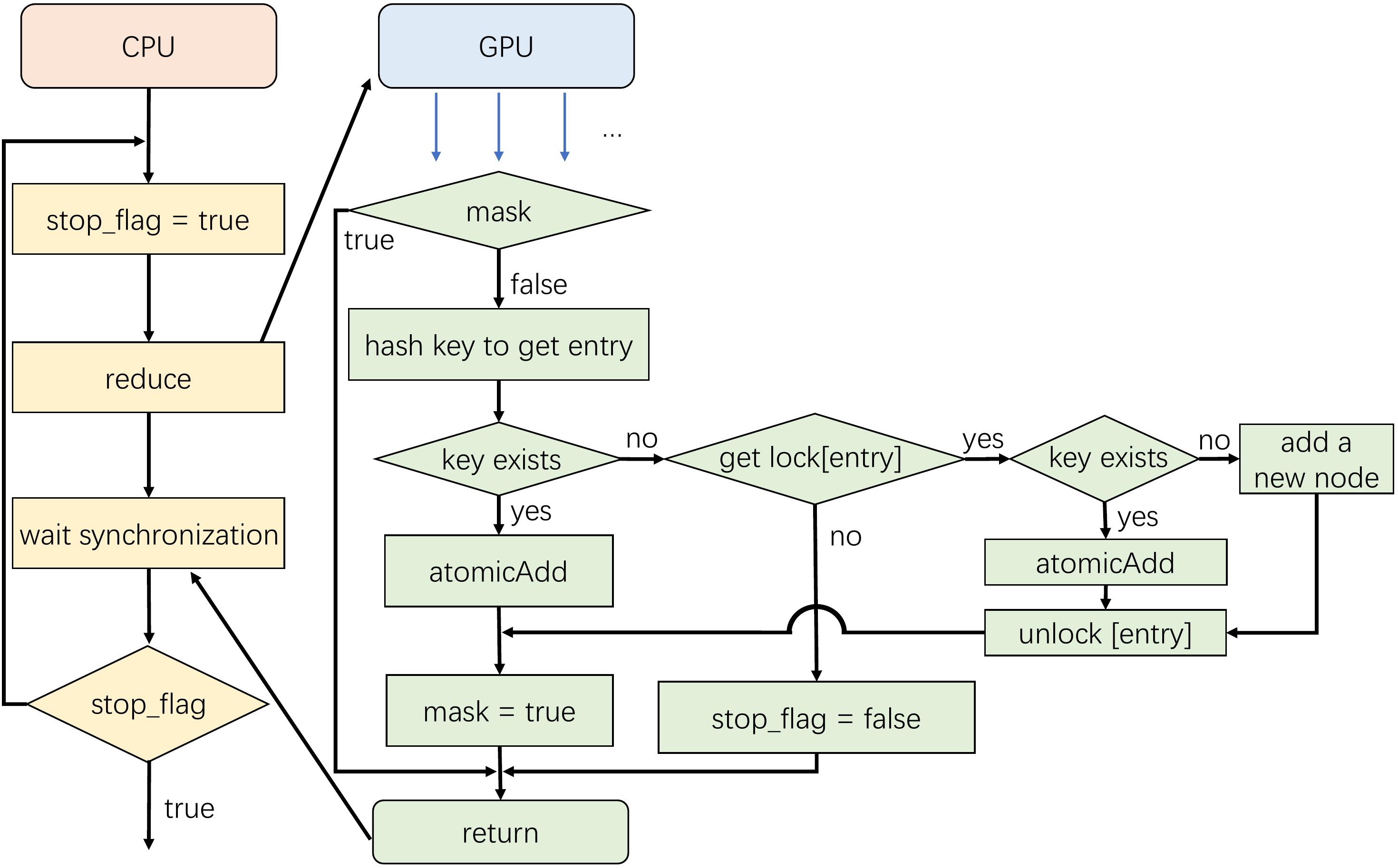}

  \caption{Phase 2: graph traversal with sequence support.}
  \label{fig:ruleprocessing}
\end{figure}

\textbf{Graph traversal phase}.
The second phase of sequence support is shown in Figure~\ref{fig:ruleprocessing}.
Similar to the first phase shown in Figure~\ref{fig:scan1},
the CPU part uses a while-loop to control the DAG traversal process.
We use \emph{sequence count}~\cite{zhang2018efficient} as an example,
which uses the hash tables described in Figure~\ref{fig:hash}.
For \emph{sequence support},
we need to reduce the intermediate results in the local tables  from the rules.
We use parallel hash tables to merge these results,
as discussed in Section~\ref{subsec:dataStructure}.
First,
we distribute each key-value pair a \emph{mask},
and each entry a \emph{lock}.
Second,
each thread is responsible for one key-value pair.
Third,
each thread needs to justify whether it is necessary to insert a key-value pair.
If not, \sysname{} returns directly;
otherwise, \sysname{} obtains the entry based on hash functions,
and then verifies if the same key already exists on this entry.
If the key exists,
\sysname{} uses atomic additions directly, and then sets the mask to \emph{true};
otherwise, \sysname{} tries to obtain the lock of the entry.
If the lock is occupied by other threads,
\sysname{} sets the stop flag to \emph{false} and returns directly;
if \sysname{} obtains the lock,
\sysname{} needs to verify whether the same key coexists.
If the same key coexists, \sysname{} uses atomic additions to avoid this issue;
otherwise,
\sysname{} obtains a new node and sets the entry accordingly.
Finally, \sysname{} unlocks the table, sets the mask to \emph{true}, and returns.
Note that the CPU part continuously launches this process until the stop flag is set to \emph{true}.

\begin{figure*}[!h]
\centering
  \begin{minipage}[t]{0.32\linewidth}
    \centering
    \subfigure[\cb{Pascal (GeForce GTX 1080).}]{
    \includegraphics[width=1\linewidth]{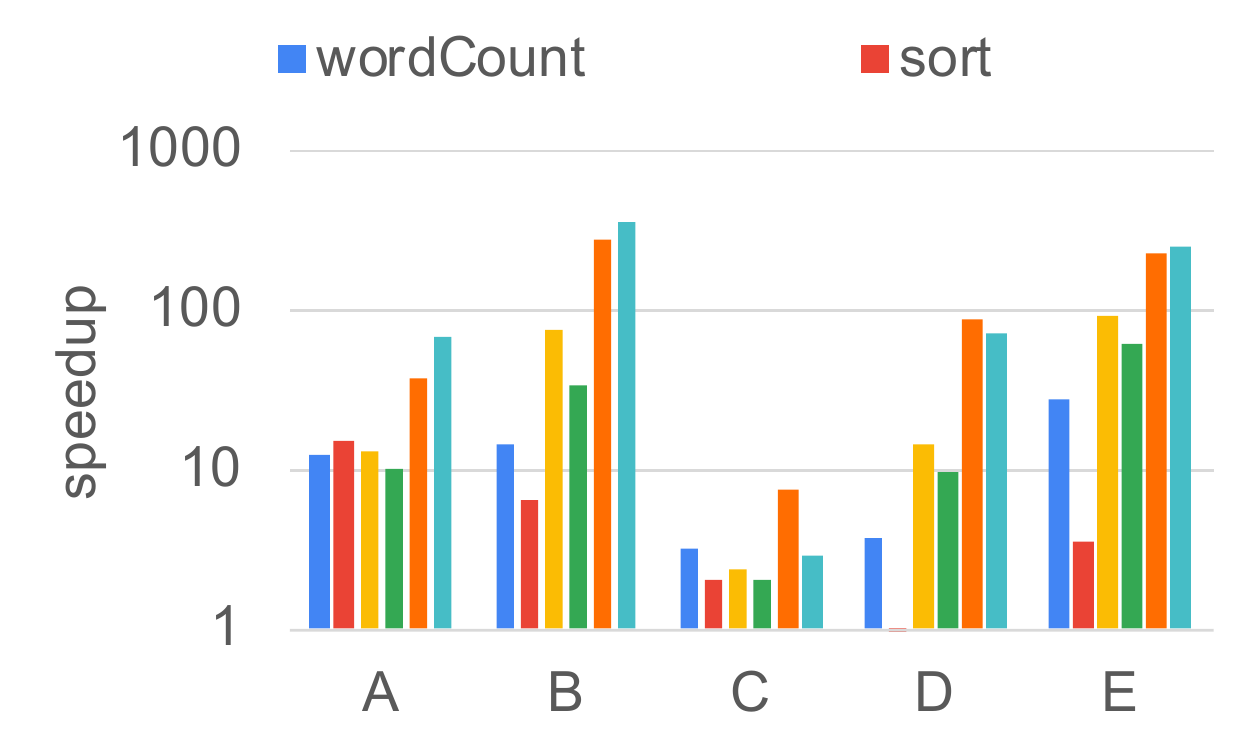}

}
  \end{minipage}
  \begin{minipage}[t]{0.32\linewidth}
    \centering
    \subfigure[\cb{Volta (Tesla V100).}]{
        \includegraphics[width=1\linewidth]{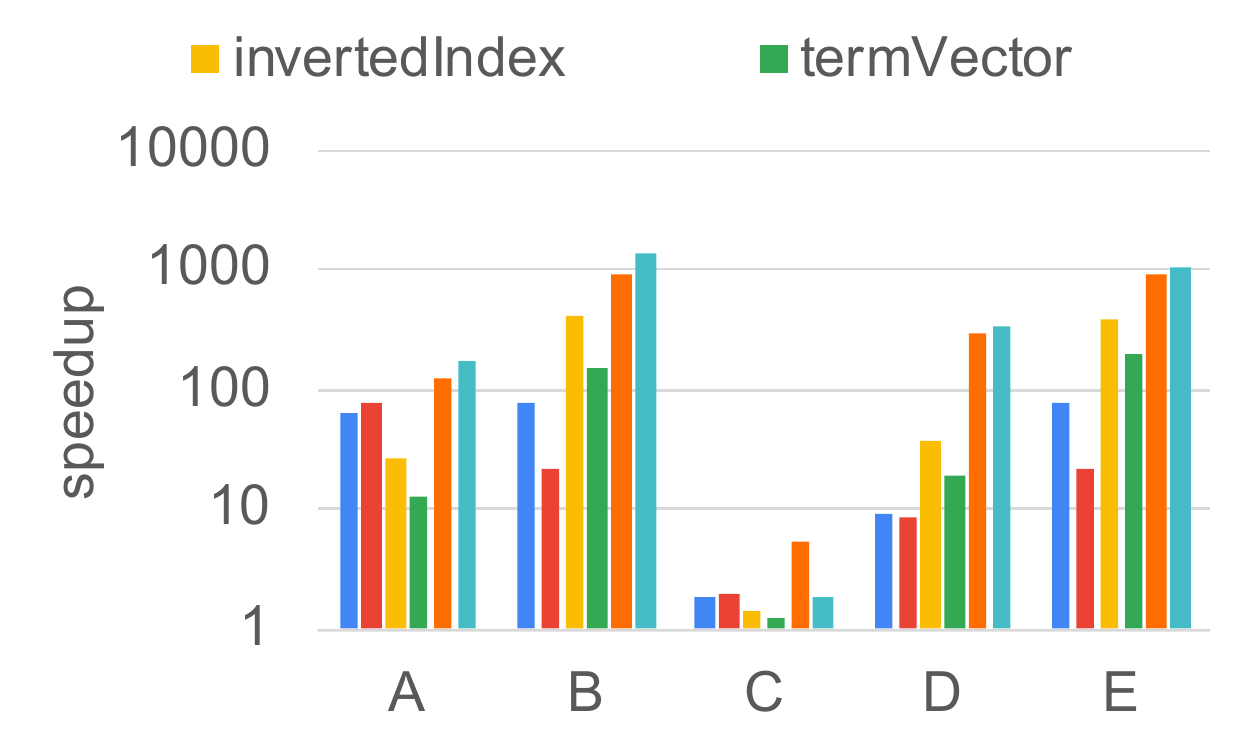}
}
  \end{minipage}
  \begin{minipage}[t]{0.32\linewidth}
    \centering
    \subfigure[\cb{Turing (GeForce RTX 2080 Ti).}]{
        \includegraphics[width=1\linewidth]{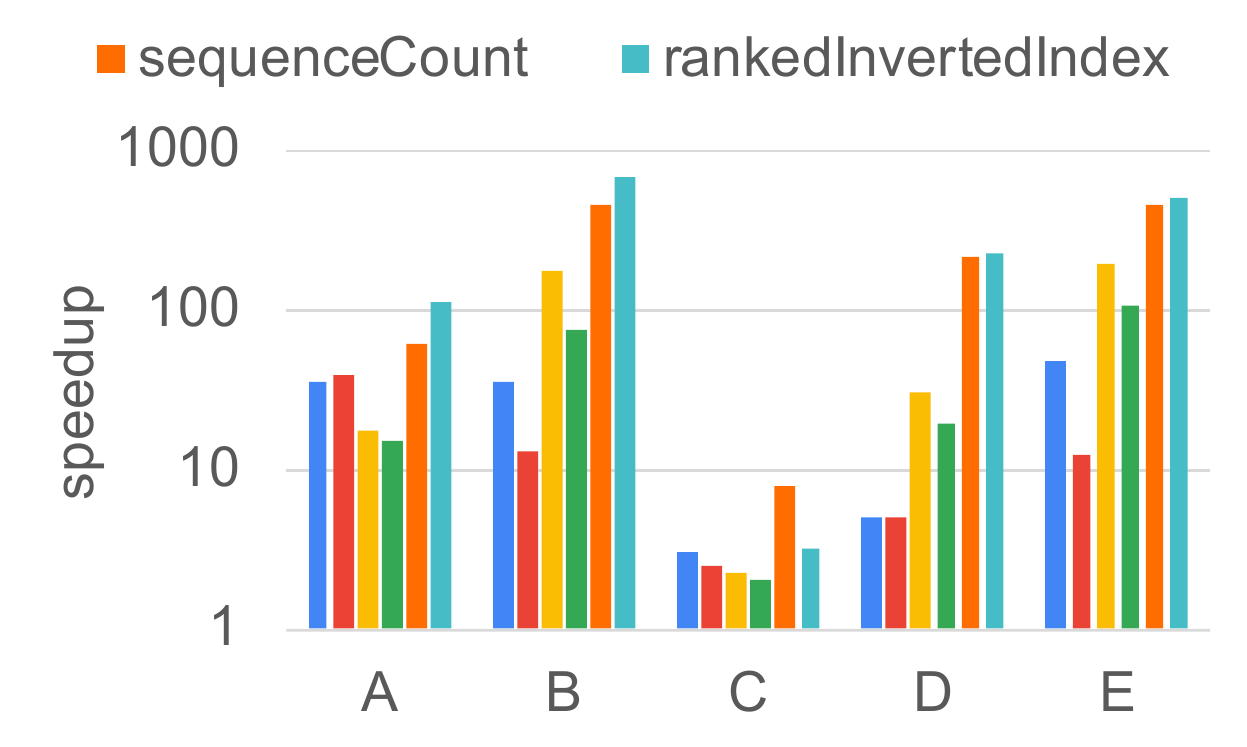}
}
  \end{minipage}

  \caption{\cb{Performance speedups.}}
  \label{fig:speedup}
\end{figure*}

\section{Implementation}
We integrate our \sysname{} into the CompressDirect (CD)~\cite{zhang2018efficient} library,
which is an implementation of TADOC.
\sysname{} in CD includes two parts:
1) the CPU part that is used to input data and program, and to handle the GPU module, and
2) the GPU part that is used for GPU-based TADOC acceleration.
We use the same interfaces as TADOC in CD,
 including \emph{word count}, \emph{sort}, \emph{inverted index}, \emph{term vector}, \emph{sequence count}, and \emph{ranked inverted index},
so users do not need to change any code in this GPU support.

\section{Evaluation}
In this section, we measure the performance of \sysname{} and compare it with TADOC~\cite{zhang2018efficient} for evaluation.

\subsection{Experimental Setup}
We show our experimental setup in this part.
\vspace{\vsp{}}

\textbf{Methodology}.
The baseline in our evaluation is TADOC~\cite{zhang2018efficient},
which is the state-of-the-art data analytics directly on compression, denoted as ``TADOC''.
Our method that enables TADOC on GPUs is denoted as ``\sysname{}''.
In our evaluation,
we measure TADOC~\cite{zhang2018efficient} performance and \sysname{} performance for comparison.
\cb{
Moreover, we assume that small datasets can be stored and processed in GPU memory directly without PCIe data transmission; large datasets are stored on disk with PCIe data transmission required to be involved in time measurement.
}

\vspace{\vsp{}}

\textbf{Platforms}.
We use three GPU platforms and a 10-node Amazon EC2 cluster in our evaluation, as shown in Table~\ref{tbl:platform}.
We evaluate \sysname{} on
 three generations of Nvidia GPUs (Pascal, Volta, and Turing micro-architectures),
 which are used to prove the adaptability of \sysname{}.
 \cb{
 Since GPU architectures are  constantly changing,
if we can achieve high performance on all these platforms,
then it is very likely that G-TADOC can achieve promising results for future GPU products.
 }
 The 10-node cluster is a Spark cluster on Amazon EC2~\cite{amazon2010amazon} for large datasets of TADOC.

\begin{table}[!h]
\caption{Platform configuration.} 

\resizebox{\linewidth}{!}{%
\begin{tabular}{ccccc}
\hline
Platform    & Pascal         & Volta          & Turing      &  10-node cluster         \\ \hline
GPU         & GTX 1080       & V100           & RTX 2080 Ti     &  NULL   \\
GPU Memory  & GDDR5X         & HBM2           & GDDR6       &   DDR3       \\
CPU         & i7-7700K       & E5-2670        & i9-9900K        & E5-2676v3      \\
OS          & Ubuntu 16.04.4 & Ubuntu 16.04.4 & Ubuntu 18.04.5  &  Ubuntu 16.04.1 \\
Compiler        & CUDA 8              & CUDA 10.1              & CUDA 11.0       & GCC 5.4.0   \\ \hline
\end{tabular}
}
\label{tbl:platform}
\end{table}

\vspace{\vsp{}}

\textbf{Datasets}.
The datasets used in our evaluation are shown in Table~\ref{table:dataset},
which include various real-world workloads.
\cb{
Datasets A, B, and C are used in~\cite{Zhang2018ics,zhang2018efficient,zhang2020enabling,zhang2020tadoc}.
}
Dataset A is  NSF Research Award Abstracts (NSFRAA) downloaded from UCI Machine Learning Repository~\cite{Lichman:2013}, and  is composed of a large number of small files.
Dataset B is a collection of four web documents downloaded from Wikipedia~\cite{wikipedia}.
Dataset C is a large Wikipedia dataset~\cite{wikipedia}.
\cb{
To increase the diversity of test data,
we add datasets D and E compared to previous works~\cite{Zhang2018ics,zhang2018efficient,zhang2020enabling,zhang2020tadoc}.
Dataset D is COVID-19 data from Yelp~\cite{yelpdataset},
and dataset E is a collection of DBLP web documents~\cite{dblpdataset}.
}
\cb{
Note that only dataset C is evaluated on the 10-node cluster.
}

\begin{table}[!h]

\centering
\caption{Datasets (``size'': original uncompressed size).}

\begin{tabular}{ccccc}
\hline

  Dataset    &  \multicolumn{1}{c}{Size}     & \multicolumn{1}{c}{File \#}  & \multicolumn{1}{c}{Rule \#}   & \multicolumn{1}{c}{Vocabulary Size}  \\\hline
A  & 580MB     & 134,631 & 2,771,880 & 1,864,902   \\
B  & 2.1GB     & 4       & 2,095,573 & 6,370,437 \\
C  & 50GB      & 109      & 57,394,616 & 99,239,057   \\
D  & 62MB      & 1      & 36,882 & 240,552   \\
E  & 2.9GB      & 1      & 8,821,630 & 23,959,913   \\
\hline 
\end{tabular}
\label{table:dataset}

\end{table}

\subsection{Performance}
In this part, we measure the speedups of \sysname{} over TADOC and show their time breakdowns.

\vspace{\vsp{}}

\textbf{Overall speedups}.
We show the speedups that \sysname{} achieves over TADOC~\cite{zhang2018efficient} in five datasets in Figure~\ref{fig:speedup}.
In detail,
Figure~\ref{fig:speedup} (a) shows the speedups on Pascal platform,
Figure~\ref{fig:speedup}~(b) shows the speedups on Volta platform,
and Figure~\ref{fig:speedup}~(c) shows the speedups on Turing platform.
We have the following observations.

First, \sysname{} achieves significant performance speedups over TADOC in all cases.
On average, \sysname{} achieves \avgSpeedup{}$\times$ speedup over TADOC.
The reason is that the GPU device for \sysname{} provides much higher computing power and bandwidth than the CPU device for TADOC.
For example,
on the Pascal platform, the theoretical peak performance of the GPU is about 185.3$\times$ over the theoretical peak performance of the CPU.
Moreover,
the bandwidth provided by the GPU memory is about 8.3$\times$ over the memory bandwidth provided by the CPUs.
The performance speedups achieved by \sysname{} further prove the effectiveness of our solutions to handle the dependencies in our parallel design for GPUs.

Second, the speedups of \sysname{} over TADOC on single nodes for processing small datasets are higher than the speedups of \sysname{} over TADOC on clusters for processing large dataset C.
The average speedup of \sysname{} over TADOC on a single node is 57.5$\times$,
while the average speedup of \sysname{} over TADOC on a ten-node cluster is 2.7$\times$.
The reason is that when processing the large dataset,
TADOC adopts coarse-grained parallelism in distributed environments to improve the data processing efficiency.
However, due to the data exchange overhead between nodes in the distributed environment of TADOC,
our \sysname{} is still more efficient than TADOC.

Third, the speedups \sysname{} achieves for \emph{sequence count} and \emph{ranked inverted index} are much higher than the speedups of the other applications in most cases.
In detail, the average speedups of \emph{sequence count} and \emph{ranked inverted index} are 111.3$\times$ and  112.0$\times$,
which are much higher than the full range average speedup.
The reason is that  \emph{sequence count} and \emph{ranked inverted index} of TADOC in~\cite{zhang2018efficient} is of low performance:
as described in~\cite{zhang2018efficient}, the performance behaviors of \emph{sequence count} and \emph{ranked inverted index}  of TADOC are close to those of the original implementations on uncompressed data without compression.
As to \sysname{},
\emph{sequence count} and \emph{ranked inverted index} reuse the partial results of duplicate data and execute in parallel on GPUs.

\vspace{\vsp{}}

\begin{figure*}[!h]

\centering
  \begin{minipage}[t]{0.49\linewidth}
    \centering
    \subfigure[\cb{Phase 1: initialization.}]{
    \includegraphics[width=0.8\linewidth]{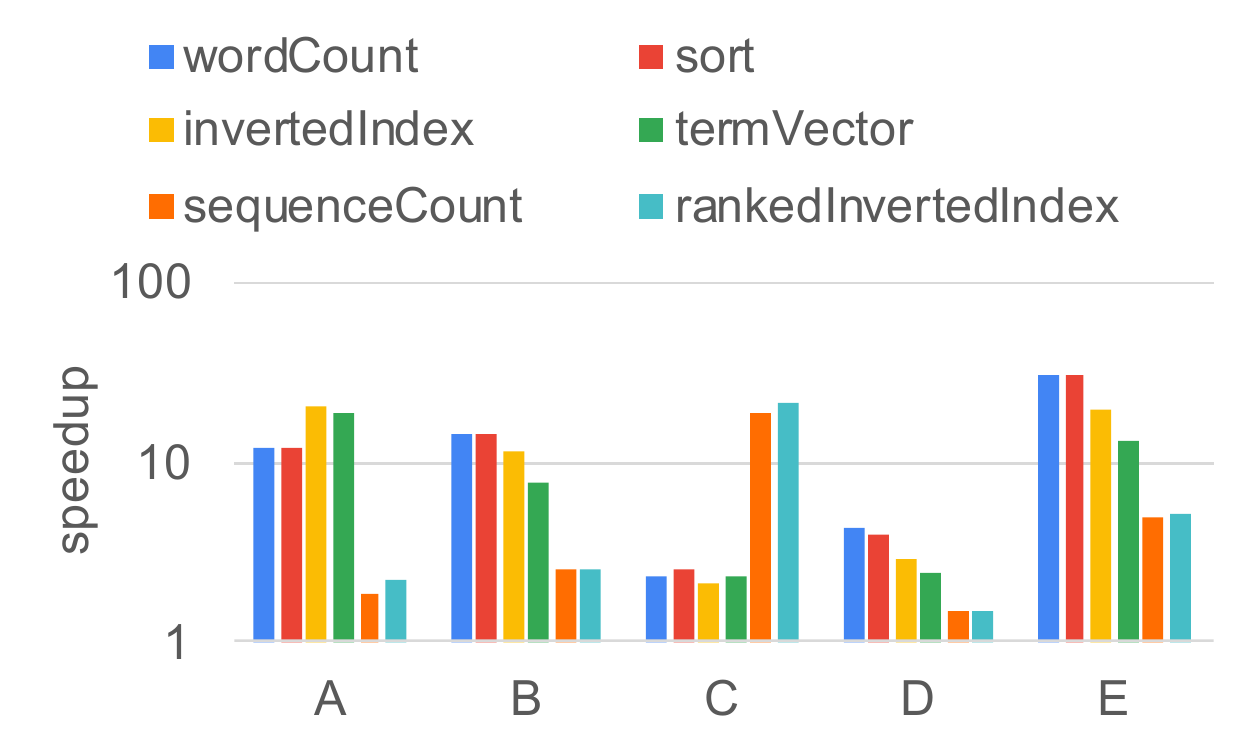}
}
  \end{minipage}
  \begin{minipage}[t]{0.49\linewidth}
    \centering
    \subfigure[\cb{Phase 2: traversal.}]{
    \includegraphics[width=0.8\linewidth]{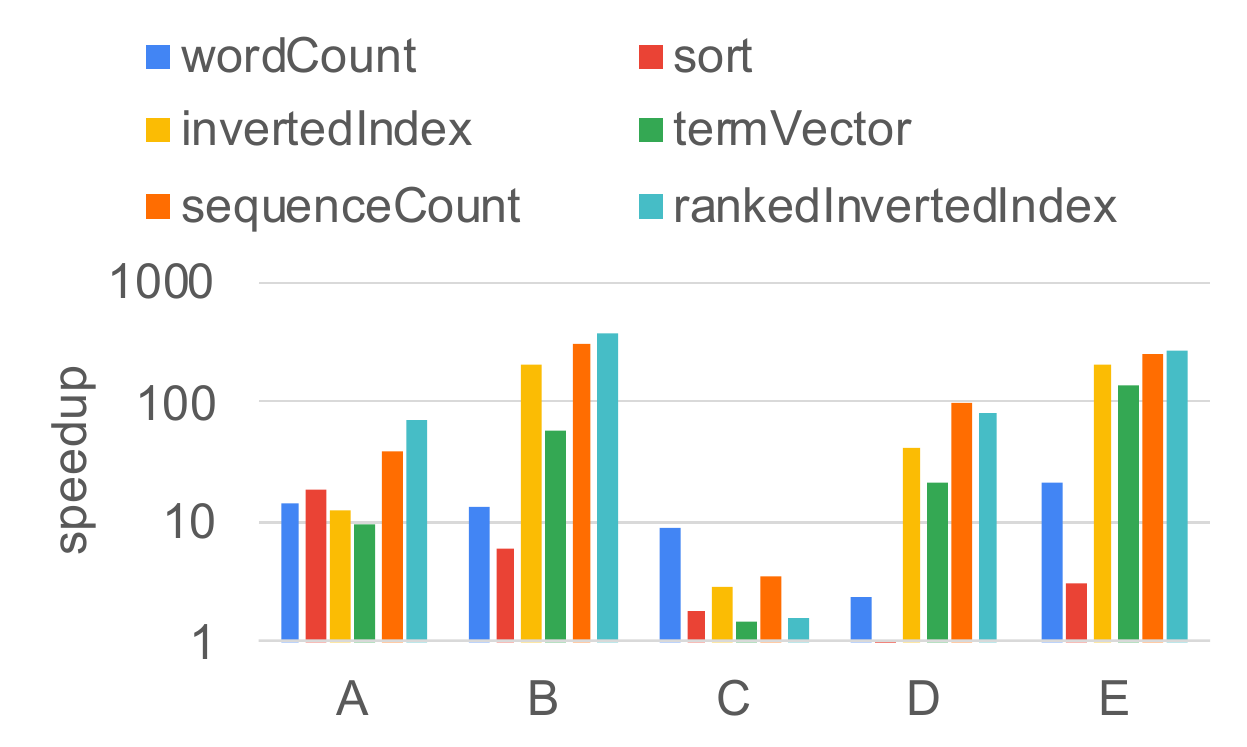}
}
  \end{minipage}

  \caption{\cb{Separate speedups for different phases.}}
  \label{fig:separatespeedup}

\end{figure*}

\subsection{Optimization Analysis}

We analyze \sysname{} acceleration in different phases and the traversal strategies on GPUs in this part.

\vspace{\vsp{}}

\textbf{Speedups in different phases}.
We show the separate speedups of \sysname{} over TADOC in different phases in Figure~\ref{fig:separatespeedup}.
First,
the average speedup in the second phase is 64.1$\times$, which implies that
most \sysname{} performance benefits come from the acceleration in the second phase of DAG traversal.
The second phase also has a relatively long execution time in TADOC~\cite{zhang2018efficient},
which provides more parallel optimization opportunities.
Second,
for the first phase,
although the execution time is relatively short,
\sysname{} still achieves clear performance speedups:
on average, \sysname{} is 9.5$\times$ faster than TADOC on CPUs.
Third,
in  large dataset C,
the speedups in the first initialization phase are extremely high,
which shows that the data structure preparation for massive large files is time-consuming in TADOC~\cite{zhang2018efficient} 
and is essential to be accelerated by \sysname{}.

\vspace{\vsp{}}

\textbf{Top-down vs. bottom-up traversals}.
We develop top-down and bottom-up traversals in \sysname{},
but the optimal traversal strategy for each application can be input dependent.
For example,
for \emph{term vector}  in dataset A,
the top-down traversal takes 14.04 seconds, but the bottom-up traversal  takes only 1.56 seconds.
In contrast,
for \emph{term vector}  in dataset B,
the bottom-up traversal takes 0.43 seconds, but the top-down  traversal  takes only 0.11 seconds.
In detail,
dataset B involves only four files.
If we traverse the DAG in a top-down strategy,
we only need to maintain a small buffer of 16 bytes in each rule indicating its file information,
and the transmission for file information in DAG traversal is also marginal.
In contrast,
for dataset A,
which involves a large number of small files,
the top-down traversal with file information would be time-consuming and drags down the overall performance.
Therefore, we should select the bottom-up traversal strategy in dataset A, and top-down strategy in dataset B.
We apply the TADOC adaptive traversal strategy selector on GPUs,
as discussed in Section~\ref{subsec:paraExe},
which can help select the optimal traversal  strategy.

\subsection{\cb{Summary of Findings}}

We summarize our findings and insight as follows.

\cb{
First, we find that GPUs are very suitable for text analytics directly on compression, but need special optimizations.
For example, \sysname{} needs fine-grained thread-level workload scheduling for GPU threads,
thread-safe data structures for parallel updates,
and head and tail structures for sequence sensitive applications.
}

\cb{
Second, the GPU platform is both cost-effective and energy-efficient,
which can be applied to a wide range of data analytics applications directly on compression,
especially in large data centers.
Experiments show that a GPU server can have much higher performance on data analytics directly on compressed data than a ten-node cluster does.
}

\cb{
Third,
although the GPU memory is limited,
our work can help put much larger content directly in GPU memory.
The frequent data transmission between the CPU and GPU drags down the performance advantages of GPUs when large workloads fail to be loaded to the GPU memory at once.
Our work  sheds light on the GPU acceleration design for such big data applications.
}

\subsection{Discussion}
We next show the importance of our paper and future work.

\vspace{\vsp{}}
\textbf{Importance of our work}.
As the first work enabling efficient GPU-based text analytics without decompression,
\sysname{} provides the insights that are of interests to a wide range of readers.
Currently,
\sysname{} involves only the applications in TADOC~\cite{zhang2018efficient},
but other data analytics tasks can all benefit from \sysname{}.
Furthermore, the series of optimizations on GPUs for TADOC can be directly applied to other advanced data analytics scenarios.

\vspace{\vsp{}}

\cb{
\textbf{Comparison with GPU-accelerated uncompressed analytics}.
In our evaluation for the six data analytics tasks with the five datasets, G-TADOC reaches \avgSpeedup{}$\times$ of the performance of the state-of-the-art TADOC on CPUs.
A common question is how the G-TADOC performance differs from the performance of GPU-accelerated uncompressed analytics.
Currently, there is no implementation about the six analytics tasks on GPUs,
so we develop efficient GPU-accelerated uncompressed analytics for comparison.
Experiments show that G-TADOC still achieves an average of 2$\times$ speedup.
}

\vspace{\vsp{}}

\cb{
\textbf{Applicability}.
G-TADOC has the same applicability as TADOC~\cite{zhang2020tadoc}.
In general, G-TADOC targets the analytics tasks that can be expressed as a DAG traversal problem,
which involves scanning the whole DAG.
}

\vspace{\vsp{}}

\textbf{How far is the performance from the optimal?}
Although G-TADOC already achieves high performance,
it still has room for performance improvement.
The reasons include 1) dependencies in DAG traversal,
2) random accesses on large memory space,
and 3) atomic operations on global buffers.
When these issues are solved,
G-TADOC can achieve at least  20\% extra performance improvement.

\vspace{\vsp{}}

\textbf{Future work}.
Currently, \sysname{} supports data analytics directly on compression on GPUs.
\cb{
This research is headed for  high-performance and efficient data analytics methods.
The future possible avenues of exploration include  architecture optimizations or multi-GPU environments,
which can further accelerate G-TADOC.
}

\section{Related Work}
As far as we know, \sysname{} is the first work that enables efficient GPU-based text analytics without decompression.
In this section, we show the related work of grammar compression, compression-based data analytics,
and GPU data analytics.

\vspace{\vsp{}}

\textbf{Grammar compression}.
There are plenty of works on grammar compression~\cite{zhang2018efficient,Zhang2018ics,zhang2020tadoc,zhang2020enabling,rytter2004grammar,charikar2005smallest,gagie2012faster,bille2015random,bille2015finger,brisaboa2019gract,ganardi2019balancing,takabatake2017space,winograd}.
The closest work to \sysname{} is TADOC,
which is the text analytics directly on compression in single-node and distributed environments~\cite{zhang2018efficient}.
TADOC extends Sequitur~\cite{nevill1996inferring,nevill1997identifying,nevill1997linear} as its compression algorithm for data analytics.
After TADOC being proposed,
Zhang \emph{et al}.~\cite{Zhang2018ics} proposed Zwift,
which the first TADOC programming framework,
including a domain specific  language,
 TADOC compiler and runtime,
and a utility library.
Then,
Zhang \emph{et al}.~\cite{zhang2020tadoc} applied TADOC as the storage to support advanced document analytics,
such as 
{\em word co-occurrence}~\cite{matsuo2004keyword,pennington2014glove},
\emph{term frequency-inverse document frequency} (TFIDF)~\cite{joachims1996probabilistic},
{\em word2vec}~\cite{rong2014word2vec,googleWord2Vec}, and {\em latent Dirichlet allocation} (LDA)~\cite{blei2003latent}.
Furthermore,
Zhang \emph{et al}.~\cite{zhang2020enabling} enabled random accesses to TADOC compressed data, and at the same time, supported \emph{insert} and \emph{append} operations.
In this work, we enable TADOC on GPUs,
which improves the performance of TADOC significantly.

\vspace{\vsp{}}

\textbf{Index compression}.
The compression-based data analytics is  an active research domain in recent years.
However, typical approaches mainly use suffix trees and indexes~\cite{navarro2016compact,agarwal2015succinct,burrows1994block,ferragina2005indexing,grossi2004indexing,ferragina2009compressed,farruggia2014bicriteria,ferragina2009bit,ferragina2009compressed,gog2014theory}.
Suffix trees are traditional representations for data compression~\cite{navarro2016compact,sadakane2007compressed} but incur huge memory usage~\cite{hon2004practical,kurtz1999reducing}.
Suffix arrays~\cite{manber1993suffix} and Burrows-Wheeler Transform~\cite{ferragina2005indexing,burrows1994block} are the development of these compression formats,
but still generate high memory consumption~\cite{hon2004practical}.
Compressed suffix arrays~\cite{grossi2003high,grossi2005compressed,sadakane2000compressed,sadakane2002succinct,sadakane2003new} and FM-indexes~\cite{wikipediaFMIndex,ferragina2000opportunistic,ferragina2001experimental,ferragina2001experimental2,ferragina2005indexing} are more efficient than the previous compression techniques.
Furthermore,
Agarwal \emph{et al}. proposed Succinct~\cite{agarwal2015succinct},
which targets queries on compressed data.
Moreover,
there are many works about inverted index compression~\cite{petri2018compact,moffat2018index,pibiri2019fast,pibiri2019techniques,pibiri2020compressed,oosterhuis2018potential,mackenzie2019compressing}.
For example,
 Petri and Moffat~\cite{petri2018compact} developed  compression tools for compressed inverted indexes.
Different from these works,
\sysname{} targets text analytics directly on compressed data on GPUs. 

\vspace{\vsp{}}

\textbf{GPU data analytics}.
GPUs have been applied to various aspects of data analytics, including structured data analytics, stream data analytics, graph analytics,
and machine learning analytics~\cite{root2016mapd,yuan2016spark,koliousis2016saber,zhang2020finestream,wang2017gunrock,pan2017multi,upadhyaya2013parallel,abadi2016tensorflow,parsecureml,sheng2020mining}.
For example, MapD (Massively Parallel Database)~\cite{root2016mapd} is a popular big data analytics platform powered by GPUs.
Most current  analytics frameworks, such as Spark, have   supported  GPUs~\cite{yuan2016spark}.
SABER~\cite{koliousis2016saber} is a stream system that schedules queries on both CPUs and GPUs,
and Zhang \emph{et al}.~\cite{zhang2020finestream} further developed FineStream,
which enables fine-grained stream analytics on CPU-GPU integrated architectures.
Gunrock~\cite{wang2017gunrock} is an efficient graph library for graph analytics on GPUs,
and for large graphs, multi-GPU graph analytics have been explored~\cite{pan2017multi}.
For machine learning data analytics,
parallel technologies have been extensively applied to various aspects, especially for deep learning applications~\cite{upadhyaya2013parallel}.
Currently, most machine learning frameworks, such as TensorFlow~\cite{abadi2016tensorflow}, support GPU.

\section{Conclusion}

In this paper,
we have presented \sysname{} enabling efficient GPU-based text analytics without decompression.
We show the challenges of parallelism, result update conflicts from multi-threads, and sequence sensitivities in developing TADOC on GPUs,
and present a series of solutions in solving these challenges.
By developing an efficient parallel execution engine with data structures and sequence support on GPUs,
\sysname{} achieves \avgSpeedup{}$\times$ speedup on average compared to state-of-the-art TADOC.

\section*{Acknowledgment}
This work is partially supported by National Natural Science Foundation of China (Grant No. U20A20226, 61802412,  and 61732014),
Beijing Natural Science Foundation (4202031 and L192027), 
  State Key Laboratory of Computer Architecture (ICT,CAS) under Grant No. CARCHA202007,
  and Beijing Academy of Artificial Intelligence (BAAI), Tsinghua University-Peking Union Medical College Hospital Initiative Scientific Research Program.
  This work is also supported by National Science Foundation (NSF) under Grants CNS-1717425, CCF-1703487, CCF-2028850, and the Department of Energy (DOE) under Grant DE-SC0013700.
  Jidong Zhai and Xiaoyong Du are the corresponding authors of this paper.

\bibliographystyle{IEEEtran}
\bibliography{IEEEabrv,ref}

\end{document}